\documentclass[letterpaper,twocolumn,10pt]{article}
\usepackage{usenix2019_v3}
\usepackage{filecontents}

\usepackage{balance}
\usepackage{soul}

\usepackage{makecell}
\usepackage{amssymb}
\usepackage{pifont}

\usepackage{graphicx}
\usepackage{tabularx}
\usepackage{booktabs}

\makeatletter
\g@addto@macro{\UrlBreaks}{\UrlOrds}
\makeatother

\usepackage[center]{subfigure}
\usepackage{comment}
\usepackage{enumerate}
\usepackage{xspace}

\usepackage{multirow}
\usepackage{soul}

\usepackage{amsthm}
\usepackage{amsfonts}
\usepackage{mathtools}
\usepackage[utf8]{inputenc}

\usepackage{enumitem}
\setlist[itemize]{noitemsep, topsep=0pt, leftmargin=10pt}

\usepackage{colortbl}
\definecolor{Gray}{gray}{0.65}
\definecolor{LightGray}{gray}{0.9}

\newcommand{\todo}[1]{\textcolor{red}{\small{TODO: #1}}}

\newcommand{\todohs}[1]{\textcolor{magenta}{\small{H: #1}}} %

\newcommand{\blue}[1]{\textcolor{black}{#1}}
\newcommand{\blau}[1]{\textcolor{black}{#1}}
\newcommand{\brown}[1]{\textcolor{brown}{#1}}
\newcommand{\red}[1]{\textcolor{red}{#1}}

\newcommand{\lsec}[1]{\label{sec:#1}}
\newcommand{\lfig}[1]{\label{fig:#1}}
\newcommand{\ltab}[1]{\label{tab:#1}}
\newcommand{\lequ}[1]{\label{eq:#1}}
\newcommand{\rsec}[1]{\S\ref{sec:#1}}  %
\newcommand{\rap}[1]{\ref{sec:#1}} %
\newcommand{\rfig}[1]{Figure~\ref{fig:#1}}
\newcommand{\rtab}[1]{Table~\ref{tab:#1}}
\newcommand{\req}[1]{Equation~\ref{eq:#1}}
\newcommand\shorten[1]{}

\newcommand{\secspacingtop}{\vspace{-10pt}}
\newcommand{\secspacingbot}{\vspace{-6pt}}

\newcommand{\oursystem}{Droplet\xspace}

\newcommand{\fakeparagraph}[1]{\vskip 2pt\noindent\textbf{#1 }}

\begin{document}

\title{\Large \bf \oursystem: Decentralized Authorization and Access Control for \\ Encrypted Data Streams}

\author{
{\rm Hossein Shafagh}\\
ETH Zurich\\
\and
{\rm Lukas Burkhalter}\\
ETH Zurich\\
\and
{\rm Sylvia Ratnasamy}\\
UC Berkeley\\
\and
{\rm Anwar Hithnawi}\\
UC Berkeley \& ETH Zurich \\
}

\date{}

\pagestyle{empty}

\maketitle

\begin{abstract}
{This paper presents \textit{\oursystem}, a decentralized data access control service. %
\oursystem enables data owners to securely and selectively share their encrypted data while guaranteeing data confidentiality 
in the presence of unauthorized parties and compromised data servers.
\oursystem's contribution lies in coupling two key ideas:
\textit{(i)}~a cryptographically-enforced access control construction for encrypted data streams which enables 
users to define fine-grained stream-specific access policies, and 
\textit{(ii)}~a decentralized authorization service that serves user-defined access policies.
In this paper, we present \oursystem's design, the reference implementation of \oursystem, 
and the experimental results of three case-study applications deployed with \oursystem:
Fitbit activity tracker, Ava health tracker, and ECOviz smart meter dashboard, demonstrating \oursystem's applicability for secure sharing of IoT streams.}
\end{abstract}

\section{Introduction}
\lsec{intro}

The growing adoption of IoT has led to an ever-increasing number of applications that collect sensitive user data. 
This growth has come with mounting concerns over data privacy.
To date, the norm has been that user data is collected and governed by application providers, e.g., Fitbit/Strava.
The problem with this status quo is that, because data lives in narrow and disjoint silos,
it severely limits a user's ability to control access to her data, extract additional value from it, or move data across applications.
This problem has led many -- from both the technical and non-technical communities -- to call for a new \emph{user-centric} model for IoT services, in which the storage of user data is decoupled from the application logic, and control over access to this data is in the hands of \emph{end-users} rather than service\linebreak providers~\cite{Amber,Sieve,Randy:local,GDP,Zachariah}.

However, if we are to realize this paradigm, we need system designs that tackle data privacy as a first-class citizen,
while ensuring users ability to securely, selectively, and flexibly grant \emph{data access} to third-party services\footnote{Note that users 
can delegate control to a third-party provider just like today -  this is permissible, just not the de-facto model.}. %
Realizing such flexible yet secure access control is key if we are to extract insightful value from user data, e.g., drive large-scale analytics from IoT data.

Such access control must ideally provide the following properties:
\textit{(i)}~strong data confidentiality and integrity, with cryptographic guarantees, 
accompanied with efficient cryptographic operations.
This is particularly essential in the context of resource-constrained IoT devices and the high volumes of \shorten{time series} data they 
generate.
\textit{(ii)}~fine-grained access control; specify who can access what temporal segment of a data stream.
\textit{(iii)}~no trusted intermediaries; systems today rely heavily on trusted intermediaries, e.g., for delegated access, rendering them trust bottlenecks.
In addition to the above, any solution must satisfy standard access control requirements, 
e.g., support for revocation and auditability.

No existing solution simultaneously provides all of the above properties.
The de-facto standard deployments today~\cite{oauth,aws-iam,google-iam, goldman-computersecurity, stamp-infsec} 
rely on trusted services (e.g., access control lists~\cite{acl}, Active Directory~\cite{desmond2008active}, OAuth~\cite{oauth})
and assume that the entity which enforces access control -- e.g., Fitbit or a storage provider -- is within  the 
data owner's trusted domain and consequently can see the data in the clear. 
However, this approach does not meet our goals of user-centric control; %
in fact, as many have argued~\cite{Digital-Immunity, ABEoauth,Pilatus,Sieve,oauthdemystified,Enigma},
this approach fails to provide even basic data privacy since the provider sees data in cleartext and 
consequently can share or sell data without user consent~\cite{guardian-thielman, healthapp}.

The %
alternative to the above approach is to rely on \emph{end-to-end} encryption~\cite{sporc, Sieve, Enigma, popa-cloudtrust, Pilatus, Talos, CryptDB}; 
where data is encrypted at the user device and stored encrypted at the storage provider; 
encryption/decryption is only executed at authorized parties and services, without disclosing any encryption keys to intermediaries.
This, however, introduces the challenge of selective sharing of encrypted data, i.e., supporting flexible access control policies.
Solutions adopted today for sharing encrypted data~\cite{filecoin, ipfs, Storj-security} fall short in
\textit{expressiveness} (i.e., allowing fine-grained access policies),
\textit{flexibility} (i.e., updates to access permissions),
and \textit{usability} (i.e., key management and revocation).
For instance, a common approach is encrypting data under each data consumer's public key; this approach suffers from hard-coded policies~\cite{Digital-Immunity, pgp}, and does not scale for high-volume and high-velocity data streams.
Moreover, in many cases, this solution is not viable, since data consumers are not necessarily known in advance, as is the case in the IoT's publish-subscribe model~\cite{hunkeler2008mqtt}.

The main question and the focus of this paper is:
\textit{how to realize a decentralized access control in a user-centric architecture?}
A solution to access control has two parts: 
\textit{(i)}~data protection (e.g., encrypting data such that a principal can only access the authorized data segment), and 
\textit{(ii)}~authorization (e.g., verifying the identity of a principal and authenticity of access permissions).

\shorten{
The canonical authorization approach in today's systems is to use a standalone authorization service that is decoupled from data protection;
i.e., a data owner registers the authorized principals with an authorization service, such as OAuth2~\cite{desmond2008active},
which serves as a trusted intermediary to issue and later verify bearer tokens for resources at a service provider.
Current authorization frameworks, besides suffering from several vulnerabilities~\cite{Sieve, oauthdemystified} 
have two key design problems that we address with this work.
First, they require users to put unlimited trust in the intermediaries running the services.
Few companies dominate this space and also learn about all services users interact.
Trusted third parties are, however, inherently prone to compromise~\cite{techcrunch-equifax}, misconduct~\cite{healthapp,uber,bbc}, and collusion/corruption~\cite{CA:fraud}.
Second, these schemes leave the enforcement of access control to the service provider.
Hence, they do not provide any assurance about data access, as they are decoupled from the underlying data protection.
Consequently, users have no guarantees that their data will not be shared against their will, nor that the sharing relationship will remain private. %
Our system, \oursystem, resembles an authorization service; similar to OAuth2, which does not suffer from these limitations.
}

In this paper, we devise a new system architecture and a crypto-based data access construction to address the above problems. 
\oursystem builds on three insights.
The first is that \emph{access control and authorization need to be co-designed} for end-to-end encrypted systems.
The second insight is that time is the natural dimension of accessing data streams. 
Hence, we design our access control with \emph{time as a prime access principle}.
The third \shorten{insight} is that there is a need for decentralized authorization services that operate \emph{without relying on trusted intermediaries}.
This is a difficult requirement, which we address with replicated state machines.
Such append-only distributed logs as underlying for example the certificate transparency~\cite{ct} %
or blockchains, provide guarantees about the existence and status of a shared state in an environment, where no single trusted intermediary is in charge and control, providing a virtual global witness to prevent equivocation~\cite{catena}.

While blockchains provide an alternative trust model, their use comes with challenges. 
Currently deployed blockchains exhibit a high overhead and low bandwidth due to their consensus protocols. %
While read operations are fast, chain-writes are inherently slow.
Hence, a key challenge is to bypass these limitations.
\blau{We design \oursystem such that blockchain operations are not on the critical path of reading and writing data;} {\em we store the absolute minimum control metadata in the blockchain} and outsource data streams and metadata to off-chain storage, by leveraging indirections.
This design minimizes the bandwidth requirements on the blockchain, and allows for lightweight clients,
which only retrieve block headers and the accompanied compact Merkle proofs.
\blue{\oursystem's authorization service leverages an existing public blockchain to maintain a replicated access control state machine.}  
This design allows any node to independently bootstrap the authorization state in a decentralized manner and check the access permissions (i.e., ensuring discoverability of access permissions without any out-of-band communication).
Access permissions are cached at the storage node for their hosted content, allowing low latency lookups of access permissions.

To realize the crypto-based access control in \oursystem; 
devices encrypt and sign their data locally.
Data owners register ownership of data streams and define privacy-preserving access permissions through \oursystem's authorization service.
Only authorized principals are cryptographically capable of accessing (i.e., decrypting) authorized data segments.
We design a novel key distribution and management construction to enable efficient key updates (i.e., succinct -- key size is independent of the granted data access range)
and fine-grained yet scalable sharing of both arbitrary temporal ranges and open-ended streams.
Our design builds on key regression and hash trees via a layered encryption technique.
\shorten{\oursystem can \shorten{(optionally)}guarantee data integrity such that even the data owner cannot later alter their data.}
In summary, \oursystem ensures data owner's sovereignty and ownership over their data, such that they maintain the ultimate power to selectively and flexibly share their data. \shorten{with desired parties.}

With a prototype implementation\footnote{\oursystem is available under \url{https://dropletchain.github.io/}}
of \oursystem,
we quantify \oursystem's overhead and compare its performance to the state-of-the-art systems.
When deploying \oursystem with Amazon's S3 as a storage layer, we experience a slowdown of only 3\% in request throughput compared to the vanilla S3. %
Moreover, we show \oursystem's potential as an authorization service for the serverless %
paradigm with an AWS Lambda-based prototype.
We show \oursystem's performance is within the range of the industry-standard protocol for authorization (OAuth2).
We also deploy \oursystem with a decentralized storage layer to give insights about its potential for the emerging decentralized storage services~\cite{ipfs, filecoin}.
With our example apps on top of \oursystem, we show that real-world applications with unaltered user-experience (i.e., perceived delay) can be developed.

In summary, our contributions are: %
\begin{itemize}
\item \oursystem, a new decentralized authorization service that enables secure sharing of encrypted data and works without trusted intermediaries. 
\item a new crypto-enforced access control construction
that provides flexible and fine-grained access control over encrypted data streams with succinct key states. 
\item a design that couples authorization with crypto-enforced access to mitigate the limitations  of current authorization services (lack of cryptographic guarantees) and end-to-end encrypted data (static policies).
\item an open-source prototype and evaluation of \oursystem showing its feasibility, %
and competitive performance.
\end{itemize}

\vspace{-5pt}
\section{\blue{\oursystem's Overview}}
\lsec{design}
\vspace{-5pt}

\oursystem's main objective is to empower users with full control (ownership) over their data while ensuring data confidentiality.
More concretely, we want to facilitate flexible and fine-grained secure sharing of encrypted data 
without ever exposing the data in the clear to any intermediaries including the storage and authorization services.
We define data ownership as having the right and control over data, wherein the owner 
can define/restrict access, restrict the scope of data utility
(e.g., sharing aggregated/homomorphica\-lly-encrypted data),  %
delegate these privileges, or give up ownership entirely without the need to rely on any trusted entities to facilitate this.
A true realization of this definition requires work on two fronts:
\textit{(i)}~privacy-preserving computation (i.e., differential privacy and secure computation) and 
\textit{(ii)}~secure and privacy-preserving access control of remotely stored data with strong confidentiality guarantees.
In this work, we focus on the latter, specifically in the context of \blue{data streams}.

\vspace{-5pt}
\subsection{\blue{\oursystem in a Nutshell}}
\vspace{-5pt}
At a high level, \oursystem is a decentralized access control system that enables users to securely and selectively share their 
data streams with principals.   
\oursystem's design marries a novel crypto-enforced access control construction
tailored for time-series data and a decentralized authorization service.
Our crypto-enforced access control construction enables users to express flexible stream access control policies (\rsec{key:management}). 
\shorten{Data is end-to-end encrypted yet can be selectively shared and accessed with our crypto-based data access construction.}The key idea behind our encryption-based access control is to serialize time series data into chunks where each chunk 
corresponds to a time segment and
is encrypted with a unique encryption key.
\shorten{This resembles a crypto-based access control that allows expressing access policies at the chunk granularity, 
e.g., minute-level.}
The challenge here becomes how to efficiently generate and manage a large number of unique encryption keys and allow 
expressing access polices with a minimum shared state that is then used to derive all decryption keys associated with the access 
policy. 
To address this specific challenge, we introduce a novel key management construction with a succinct key state, i.e., the key size does not grow with the temporal range of shared data~(\rsec{key:management}).
Although crypto-based access control is powerful, it is not sufficient by itself, as it does not adequately handle authorization and revocation.
To address this issue, we introduce a decentralized authorization service (\rsec{design:blockchain}) that interplays with our 
crypto-based access control construction.

Consequently, data owners are not required to exchange any encryption keys directly with data consumers.
Our decentralized authorization, in its essence, is similar to OAuth2.
However,  we realize the access control state machine on top of an existing blockchain~(\rsec{design:virtualchain}), 
and eliminate the need for trust intermediates on which OAuth2 realizations heavily depend.
The access control state machine assembles the current global state (i.e., access permissions and data ownership) through 
embedded private state transitions.

\shorten{
To realize \oursystem, we need to address the fundamental challenge of securing times-series data with an adequate access control mechanism.
Secure file storage solutions fall short in providing the adequate query interface for time-series data and
encrypted databases are an overkill for such data~\cite{shafer2013specialized}
(e.g., no need for concurrency or recovery due to single writer property).
Key-regression~\cite{fu:keyregression, Bolt} provides a low overhead key management scheme.
However, it does not support fine-grained access control expressiveness and leaves out the challenge of secure key distribution.
}

\vspace{-5pt}
\subsection{\blue{Security Model}}
\vspace{-5pt}
\lsec{sec-model}

\fakeparagraph{Threat model.}
\textit{(i)~Data storage:} \shorten{the threat model addressed by \oursystem consists of \brown{an adversary},} 
we consider an adversary who is interested in learning about users' data. %
Our threat model covers malicious storage nodes, potential real-world security vulnerabilities leading to data leakages, 
and also external adversaries who gain access to data as a result of system compromise. 
\shorten{Moreover, an adversary can launch a data scraping attack against storage nodes.}
\textit{(ii)~Access Permissions State:}
an adversary may access and bootstrap the access control state machine, but it cannot alter or learn sensitive information about the access permissions (e.g., sharing relationships or keying material).
For an adversary to alter the access permission states, it needs to break the security of the underlying blockchain.
The standard blockchain threat model assumes that an \blue{adversary cannot control a large percentage of nodes in the network, for the blockchain to be considered secure.} 
The actual ratio depends on the deployed consensus protocol by the underlying blockchain.
For instance, given $n$ blockchain nodes and $f$ adversary nodes,
a ratio of $n = 2f +1$ for Nakamoto-style consensus mechanisms~\cite{nakamoto2008bitcoin} or $n = 3f +1$ for PBFT consensus mechanisms~\cite{bano2017consensus} is required for the honest majority.

 \begin{figure}[t]
	\begin{center}
	\includegraphics[width=1\columnwidth]{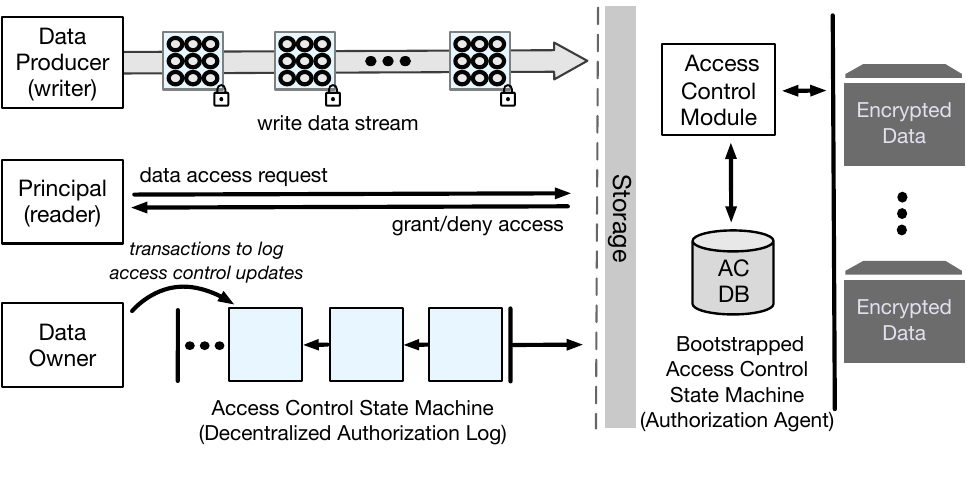}
	\vspace{-14pt}
	\caption{Abstract protocol flow.
	Data is E2E encrypted with encryption-based access control.
	The data owner stores access permission updates in the decentralized authorization log.
	The storage service validates access requests based on the access permissions from the access control state machine.
	}
	\vspace{-11pt}
	\lfig{system_overview}	
	\end{center}
\end{figure}

\fakeparagraph{Guarantees.}
\oursystem embodies a decentralized encryption-based access control mechanism that enables secure and selective 
access to stream data within the above-discussed threat model.
Data is encrypted at the client-side, and keys are at no time disclosed to intermediaries, i.e., storage and authorization services, guaranteeing \blau{data confidentiality, integrity, and authenticity.}
Decryption keys are only shared with authorized parties via a blockchain-based indirection, ensuring \textit{asynchronicity},
i.e., keys are established without requiring participants to be online at the same time.
In case decryption keys are compromised, \oursystem guarantees that only the user's data stream segment associated with the key is disclosed,
and the compromised keys cannot be used to disclose past or future data beyond the temporal segment associated with the key.
Data partitions are signed, allowing parties without decryption keys to verify data authenticity and integrity.
\oursystem enables checking the freshness of data, and
it provides data immutability optionally via an authenticated data structure anchored in the blockchain,
such that even the data owner can no longer modify past data.
\oursystem cryptographically prevents evicted users from accessing future data.
Though evicted users may have already cached past data, they are, however, prevented from future access.
\oursystem encodes user-defined access permissions in the blockchain, eliminating trusted intermediaries and assuring collusion-resistance and auditability.
\shorten{Even malicious institutions cannot illegitimately modify access permissions.}
Moreover, we employ privacy-preserving access permissions, preventing an observer from learning the sharing parties' identities.
\oursystem does not protect against denial-of-service attacks, nor does it hide access patterns.
It could be extended with ORAM techniques to hide access patterns~\cite{OblivP2P, stefanov2013oblivistore}.
Cryptographic techniques alone are not sufficient to prevent a malicious storage provider from denial-of-service or deconstruction of data.
Hence, adequate replication strategies on multiple providers are necessary to ensure the preservation and availability of data.
\blue{In \rsec{sec-guarantees}, we discuss the security guarantees in more detail.}

\fakeparagraph{Assumptions.}
In \oursystem, we make the following assumptions.
We assume the storage nodes to be available.
This is a valid assumption since storage nodes can face financial (and potentially legal) consequences upon detection of misbehavior.
\oursystem guarantees data confidentiality even if malicious storage nodes hand over data illegitimately, as data is end-to-end encrypted.
We assume the adversaries to be subject to the standard cryptographic hardness
and the underlying blockchain to be secure,~i.e., similar to previous work~\cite{Blockstack, catena, MPC-Bitcoin, fairBitcoin}, we assume transactions are \blue{append-only, ordered, and immutable after a confirmation period and the blockchain to be highly available.}
We assume users store their keys securely 
and that key recovery techniques are deployed (we discuss in \rsec{discussion} potential recovery techniques, such as Shamir's secret sharing).
We assume data producers to report correct data and to perform data serialization and encryption correctly.
We assume there is a financial agreement between the storage provider and data owner to provide persistent storage,
which can \shorten{optionally}also
be facilitated through the cryptocurrency feature of the underlying blockchain.

\vspace{-5pt}
\subsection{\blue{Architecture}}
\vspace{-5pt}
As illustrated in \rfig{system_overview}, our design considers four actors and three system components:
\textbf{data owner} is someone who owns a set of devices (e.g., wearables, appliances, services) which produce time-series 
data, i.e., {\bf data producers}.
In an industrial setting, the data owner can be an organization that owns a swarm of IoT devices.
The generated data is stored on storage services, and
data owners can decide to selectively expose their data to data consumers (i.e., {\bf principals}) who can produce an added 
value from the data
(e.g., fuse several streams for prediction tasks).
\blue{Data is end-to-end encrypted at the data producer, and
each principal computes the corresponding decryption keys locally based on an encrypted authorization token (i.e., embodies the access policy state) shared through \textit{\oursystem's decentralized authorization log}.}
Data owner, data producer, and data consumer run \textit{Droplet's client library}, which covers the tasks of data serialization, enc/decryption, key management, 
and setting/viewing access permissions.
\blue{Moreover, end-user applications (e.g., Fitbit/Strava) interact directly with Droplet's client API to facilitate sharing through \oursystem.}
The \textbf{storage} node is in charge of storing encrypted data and providing access to principals as defined by the data owner.
The storage node grants or denies access requests via \oursystem \shorten{decentralized authorization log}, i.e., in accordance with 
user-defined access permissions. 
Access permissions are cryptographically bound to a specific principal's identity (public key).\shorten{, who can claim ownership via the corresponding private key.}
The storage node can take various forms, such as edge, decentralized (e.g., a node in a p2p storage 
service~\cite{ipfs}), 
or cloud storage (e.g., Amazon's S3).
The storage node runs \oursystem's storage engine and can additionally run \textit{\oursystem's authorization agent} to handle access requests locally.
\textit{\oursystem's authorization agent} bootstraps its state from the \textit{decentralized authorization log}.
As a matter of fact, anyone can run \textit{\oursystem's authorization agent} to either expose it as a service or to monitor the state of relevant access permissions. 
Note that \oursystem's decentralized authorization agents are stateless and
cache relevant access permissions for fast lookup, e.g., maintaining access permissions of resources stored by the storage node.
\shorten{We now introduce \shorten{and discuss} different components of \oursystem.}

\shorten{We begin with a simplified description of our system components and gradually converge to the full system design.}

\blue{
\oursystem is, in essence, a new decentralized access control system that is materialized by coupling a new encryption-based access control scheme and a decentralized authorization service.
In the following, we elaborate on our encryption-based data access construction.
As the backbone of our encryption-based data access, we present the design of an efficient key-management construction.
Afterward, we discuss \oursystem's decentralized authorization service.}

\vspace{-5pt}
\section{\blue{Encryption for Access Control}}
\lsec{key:management}
\vspace{-5pt}

\fakeparagraph{Goals.}
With our crypto-enforced data access construction, we pursue a design that fulfills the following goals:
\textit{(i)~Flexible sharing abstractions}:~support of the three common types of sharing modalities desired for time-series data, varying 
based on the role and purpose of the data consumer;
\textit{(a)}~subscription, where the data consumer is granted continuous access to the data stream as it is generated, either 
temporarily or until revoked, 
(e.g., a visualization app rendering an overview of the user's daily activity based on wearable data),
\textit{(b)}~sharing arbitrary intervals of past data 
(e.g., a practitioner app accessing and analyzing user's health data during past pregnancy),
and \textit{(c)}~a combination of \textit{i} and \textit{ii}.
\textit{(ii)~Efficiency:} computationally efficient crypto primitives to adhere to the constraint resources of IoT devices,
\textit{(iii)~Scalability:}  to cope with the velocity and large volume of time-series data. %

\fakeparagraph{Gist:} 
A key aspect of our construction is tied to the observation that time-series data streams are continuous.
Hence, we introduce time-encoded key-streams which map keys to temporal segments of the data stream, such that access to the data stream can  be restricted by only sharing the corresponding range in the keystream with a principal.
Based on the access policy, the principal gains access to the necessary decryption keys via an access token.
Access tokens are encrypted with the principal's public key (hybrid encryption).
To enable sharing without enumerating all the keys and expressing stream access policies in a succinct shared state, we design a key derivation construction 
that synthesizes the concepts underlying hash trees and dual-key regression.

\secspacingtop
\subsection{\blue{Encryption-based Access Control}}
\secspacingbot
Each data chunk of a data stream is encrypted under a random symmetric key derived from our key derivation construction.
Keys are rotated for each chunk %
permitting access permissions at the chunk level.
This allows for flexible access policies for individual data consumers without the need for data re-encryption or introducing redundant data.
The design of our key derivation construction in its core builds on hash trees~\cite{briscoe1999marks} and key 
regression~\cite{fu:keyregression}
to enable expressing \emph{stream-specific access policies} and \emph{efficient management of encryption keys.}
\oursystem supports computing a large segment of keys from a single shared state instead of sharing individual keys.

We now give a brief background on hash trees and key regression and their role in our encryption-based access control construction.
We elaborate why these two components alone fall short in meeting our design requirements and describe how 
we leverage them to create our hybrid key management construction.
We formalize the security guarantees of our key management in \rap{proof}.

 \begin{figure}[t]
	\begin{center}
	\includegraphics[width=.99\columnwidth]{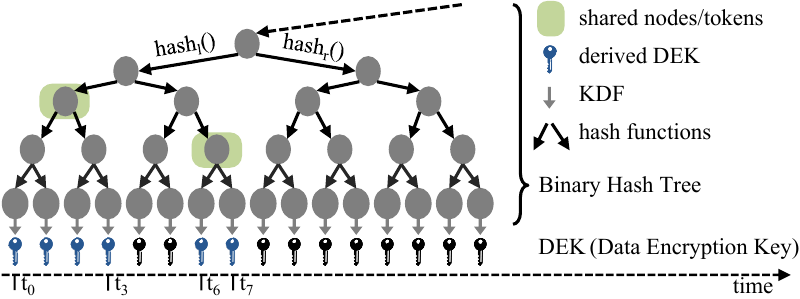}
	\vspace{-5pt}
	\caption{\oursystem's key generation. %
	Data Encryption Keys (DEKs) are managed through the hash tree, allowing efficient sharing of arbitrary intervals.
	An access policy contains several shared nodes as authorization tokens.
	}
	\vspace{-11pt}
	\lfig{tree-sharing}	
	\end{center}
\end{figure}

\fakeparagraph{Binary Hash Tree (BHT).}
A BHT~\cite{briscoe1999marks} is a balanced binary tree, built top-down from a secret random seed as the root;
using two cryptographic hash functions for the left and right child nodes, i.e., hash$_l$() and hash$_r$(), respectively.
Initially the hash functions are applied to the root node.
This procedure is applied recursively until the desired depth $h$ in the tree is reached, as depicted in \rfig{tree-sharing}.
The leaf nodes represent the keystream $\{k_0, k_1, k_2, ..., k_{2^h-1}\}$.
We select a large $h$ such that the keystream is virtually infinite.

We encrypt each data chunk of the data stream with a unique key derived from the BHT. %
With this construction users can efficiently share any arbitrary time interval of their stream; 
by just sharing the inner nodes in the BHT necessary to compute the corresponding keys.
For instance, in~\rfig{tree-sharing}, given the two highlighted inner nodes a data consumer is granted access to two disjoint 
intervals  $t_{[0-3]}$ and $t_{[6-7]}$, and can compute the corresponding decryption keys.
While consistent with our efficiency and low overhead requirements, this BHT-based construction lacks support for sharing in subscription mode, where data 
consumers have continuous access to data streams.
Realizing this mode of sharing with BHT requires maintaining and sharing a growing state per individual data consumer.

\fakeparagraph{Key Regression.}
Key regression~\cite{fu:keyregression} is a hash-chain based construction that enables sharing a large number of keys by only sharing a single state.
Given a single hash token, one can derive all previous keys by applying the hash function successively,
i.e., given key $K_t$ in time $t$ one can compute all keys until the initial key $K_0$, i.e., $\forall_{i\in [0..t]} K_i$.
However, no future keys can be computed (forward-secrecy).
This is not always desirable, as key regression enables sharing of all keys from the beginning until \emph{current time}
(all-or-none principle).

\fakeparagraph{Dual-Key Regression.}
To overcome the all-or-none limitation of key regression, we design a hash chain construction that enables sharing with a defined lower time bound, e.g., access to data of a particular stream from \emph{Nov'18} till revoked.
To realize this, we extend key regression with an additional hash chain in the reverse order, to cryptographically enforce both boundaries of the shared interval (\rfig{keyregression}). 
In simple key regression, hash tokens are consumed in the reverse order of chain generation as input to a key derivation function to derive the current key.
Due to the pre-image resistance property of hash functions, it is computationally hard to compute future tokens and hence future keys. 
However, the reverse can be computed efficiently. 
We leverage this property of hash chains for defining the beginning of an interval through a secondary hash chain in the reverse order, as depicted in \rfig{keyregression}.
In the \textit{dual-key regression}, the Key Derivation Function (KDF) takes a second token $h'_i$: $KDF(h_i || h'_i$)~=~$K_i$,
with $h'_i$ from the secondary hash chain (\rfig{keyregression}).
For instance, to share a data stream from time $t_i$ to $t_j$, the user provides the tokens $h'_i$ and $h_j$.
Since it is infeasible to compute $h_{j+1}$, no key posterior to $k_j$ can be computed.
Conversely, since it is infeasible to compute $h'_{i-1}$, no key prior to $k_i$ can be computed.
With access to the two hash tokens ($h_j$, $h'_i$), indicating the beginning and end of the shared interval, one can compute all the encryption keys within this interval.
We formalize and prove the security guarantees of dual-key regression in \rap{dkr}.

\begin{figure}[t]
	\begin{center}
	\includegraphics[width=.99\columnwidth]{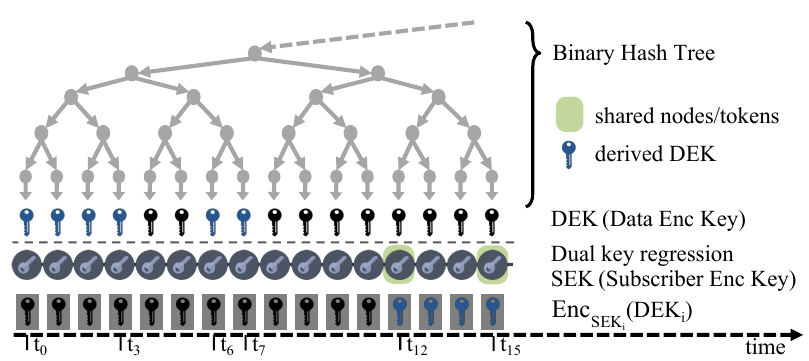}
	\vspace{-9pt}
	\caption{\oursystem's hybrid key management supports sharing of arbitrary intervals (hash tree) and subscriptions (dual-key regression).
	Given the opening and end tokens (dual-key regression), one computes the interval Data Encryption Keys. \shorten{(DEKs) locally.}
	}
	\vspace{-14pt}
	\lfig{advancedkeymanagement}	
	\end{center}
\end{figure}

\secspacingtop
\subsection{\blue{\oursystem's Key Management}}
\secspacingbot
We now discuss how our design compounds 
\textit{dual-key regression} and  BHT via a layered encryption technique to enable\shorten{our desired} stream sharing abstractions.
Dual-key regression resembles a linear chain of keys, \blue{where for a given state,} i.e., \textit{beginning} and \textit{end} tokens, one can compute all the keys in between.\linebreak
Conceptually, we exploit the hash tree to allow arbitrary sharing of intervals and the dual-key regression to support sharing in subscription mode.

The layered encryption consists of two steps:
\textit{(i)}~the hash tree delivers time-encoded data encryption keys $DEK_i$, which we use to encrypt data generated during the time epoch~$i$.
\textit{(ii)}~the dual-key regression also delivers time-encoded subscriber encryption keys $SEK_i$ for the epoch~$i$.
We use $SEK_i$ to encapsulate the corresponding data encryption key:  $ENC_{SEK_i}(DEK_i)$.
For fast access, each encrypted data chunk holds the encapsulated $DEK$.
With this construction, we can give access to data encryption keys either via the hash tree (arbitrary intervals) or dual-key regression (subscription), as depicted in~\rfig{advancedkeymanagement}.
To a subscriber, $DEK$s appear as random encryption keys.
For principals with access to past data, $DEK$s are the leaf nodes of the BHT which they locally compute based on the shared inner nodes (e.g., root nodes of the corresponding subtrees).
Note that a principal can be granted access in both modes simultaneously,
 as shown in the example of~\rfig{advancedkeymanagement}.
In this example, the data owner has granted the principal access to the intervals $t_{[0-3]}$ and $t_{[6-7]}$, %
which is realized through the hash tree.
Also, the principal is granted a subscription from $t_{12}$ which is realized over dual-key regression.
We describe next how to handle long key chains efficiently and in constant space.

 \begin{figure}[t]
	\begin{center}
	\includegraphics[width=1\columnwidth]{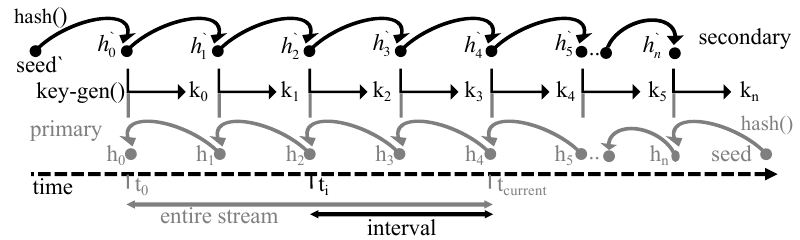}
		\vspace{-15pt}
	\caption{The dual-key regression supports time-bounded sharing via a secondary hash chain.
	The gray elements depict the standard key regression mechanism:
	Given current $k_c$, one can compute all keys up to $k_0$.
	Our construction allows the sharing of keys for an interval via a secondary hash chain.
	}
		\vspace{-12pt}
	\lfig{keyregression}	
	\end{center}
\end{figure}

\fakeparagraph{Key Distribution.}
An important aspect to address in crypto-based access control schemes is how to distribute 
keys efficiently. In \oursystem, this is especially tricky for the subscription mode, where new 
data chunks arrive continuously, and each one is encrypted with a new key.
We now describe our key distribution mechanism and refer to \rsec{design:blockchain} for insights on obtaining the keying material over the decentralized authorization service.\shorten{storage network.}
When a new data consumer is added, an authorization token encapsulating the defined access policies is issued which contains either: 
\textit(i)~the state to compute decryption keys for past data intervals (i.e., inner nodes of the hash tree) or 
\textit(ii)~in case of sharing in the subscription mode the hash token for the beginning of the interval $h'_i$ (i.e., dual-key regression).
For the subscription mode the challenge is to give the active set of subscribers continuous access to the latest token (i.e., $h_{t}$ from the main chain),
such that they can compute the current decryption key.
If we were to encrypt the current hash token for each subscriber individually, this would incur communication/computation overheads in $O(s)$, given $s$ subscribers.

To reduce this overhead, we distribute the latest dual-key regression token $h_t$ within a digitally signed and encrypted lockbox.
Authorized subscribers obtain the long-term \textit{distribution key} $KD$ to open the lockbox.
This approach is more efficient than resorting to per subscriber encryption.
When sharing access to a data stream, we share the distribution key encrypted for the new subscriber 
through the authorization service~(\rsec{design:blockchain}). 
While \textit{data encryption keys} and hence dual-key regression tokens are frequently updated at a defined interval,
the distribution key is only updated after an access revocation event, as detailed next.

A subscriber decrypts the current data encryption key $ENC_{SEK_{j+1}}(DEK_{j+1})$ given the current token $h_{j+1}$ and the opening token $h'_i$ as: 
\vspace{-7pt}
\begin{equation}
KDF(h_{j+1} || h'_{j+1}) = SEK_{j+1}, \text{with~H}^{(j-i+1)}(h'_i) = h'_{j+1} \vspace{-2pt}
\end{equation}
with $H$ as a hash function.
The secondary token is stored along the long-term per principal key information (\rsec{design:blockchain}). %

\fakeparagraph{Revocation.}
To revoke data stream access, the data owner updates the distribution key (i.e., crypto-based access) and
issues a state update transaction (i.e., authorization) to evict the revoked service.
The transaction includes a new distribution key $KD'$ contained in the encrypted key information per subscriber.
Hereafter, the new data encryption key is only available to the remaining authorized subscribers, protected with the new distribution key.
\blue{The transaction confirmation time of the underlying blockchain determines the delay until \oursystem's authorization state machine is updated.
The end-to-end encryption, however, prevents revoked users from accessing new data instantly, due to the preceding key rotation.}

With the newly issued transaction, the global access permission state is updated (\rsec{design:blockchain}).
\oursystem cryptographically prevents any future access to new data by the evicted subscriber.
Any future access requests by the evicted subscriber to old data are declined during authorization.

\fakeparagraph{Compact Hash Chains.}
Our key management, %
specifically dual-key regression,  relies heavily on hash chains.
The underlying chains can grow quickly due to frequent key updates.
\blue{Given the memory-constraints of IoT devices, we revert to a combination of re-computing on-demand and storing a segment of the hash chain in memory, to achieve fast and efficient key rotations.}
We leverage hierarchical hash chains~\cite{hu2005efficient} which maintain the same security features as traditional hash chains but reduce the worst-case compute time %
to $O(\sqrt{n})$. %
In our evaluation in \rsec{eval:data}, we show how compact chains allow for a two-orders of magnitude key rotation speed-up.

\vspace{-10pt}
\section{\blue{Decentralized Authorization Service}}
\lsec{design:blockchain}
\vspace{-8pt}

\blue{So far, we have covered \oursystem's encryption-based access control mechanism.}
Now we describe \oursystem's authorization service which handles and manages access permissions. 
At a high level, through \oursystem's API, users can view their data streams, the associated sharing policies, and storage 
information, and can set/edit access permissions accordingly.
Similar to today's authorization frameworks, e.g., OAuth2, our authorization service acts on behalf of users, forgoing direct interaction of individual services with the data owner. 
Storage providers query \oursystem's authorization agent directly to validate access requests. 
Moreover, principals query the authorization agent to retrieve authorization tokens. \shorten{for computing decryption keys.}
\blue{The authorization agent falls under the same trust assumptions as the storage node, which enforces the authorization verdict.
This means that the storage node can act maliciously, i.e., bypass the agent's authorization verdict, 
and hand out data to unauthorized parties.
Similarly, an authorization agent can also act maliciously.
However, due to \oursystem's end-to-end encryption, these violations do not compromise data confidentiality (\rsec{design:privacy}).}

In our design,
we employ a tamper-proof decentralized authorization log to enable anyone to bootstrap and presume the role of 
\textit{authorization agent} and serve access permission lookups in a decentralized and verifiable manner.
We realize the authorization log using a publicly verifiable blockchain to maintain an accountable distributed 
access control system \textit{without a central trusted entity}.
This allows us to move away from a single authorization server issuing and verifying access tokens, to where any resource 
owner can issue access permissions and any node can verify it. 
\shorten{\red{\brown{The reasons why we employ the blockchain technology to realize our decentralized authorization log are manifold:}
\textit{(i)}~resilience against vulnerabilities of trusted intermediaries,   %
\textit{(ii)}~distributed identity management, specifically of relevance to the IoT, 
\textit{(iii)}~transparent audibility of access permissions by authorized parties,
\textit{(iv)}~potential of nano-payments for storage services and data market.
\oursystem embeds data stream ownership and access permissions in the blockchain transactions.} \todohs{remove?}}
We now describe the owner-device pairing, blockchain-embedded access permissions, and how we protect the privacy of principals.

\fakeparagraph{Owner-Device Pairing.}
The blockchain ecosystem relies on public key cryptography for identification and authentication of the involved principals.
The hash digest of the public key serves as a unique pseudo-identity in the network.
We leverage this feature to allow IoT devices to securely and autonomously interact with the storage service. %
This way we overcome the hurdle of passwords and rely on public-key crypto for authentication and authorization.
During the bootstrap phase of a new device, it creates a pair of public-private keys locally, where the private key is stored securely and never leaves the device.
Through an initial two-way multisignature registration transaction on the blockchain, \oursystem allows the binding of  \blue{IoT devices ($PK_{device_{i}}$, $SK_{device_{i}}$) to the owner ($PK_{Owner}$, $SK_{Owner}$).}
Henceforth, the owner can set access permissions \blue{(via the private key $SK_{Owner}$)
and the IoT devices are permitted to store data (via the private key $SK_{device_{i}}$)} securely.
The necessary keying material for encryption (\rsec{key:management}) on the data producer is also exchanged during the initial phase.
Note that the data owner's \blue{private} key is powerful in that it sets/updates access permissions. 
\blue{\oursystem assumes a data owner private key management scheme to be in place (e.g., Human-Memorable Password-Protected Secret Sharing, backed with hardware security modules or multiple trusted devices~\cite{bagherzandi2011password, jarecki2014round}),
and a key recovery mechanism to be employed for handling a potential key loss (see \rsec{discussion}).}

\blue{In the event of device decommissioning, the new owner must issue a new multisignature device-binding transaction, to gain ownership of future data produced by the same device.}
Note that there is no need for the IoT device to interact with the blockchain \shorten{network}directly.
The owner creates the raw multisignature registration transaction and uses an out-of-band channel
(e.g., Bluetooth Low Energy) to get the device's signature.
After adding her signature, she broadcasts the register transaction to the network. %
\blue{Note that the channel between the IoT device and owner must be secure, otherwise we risk disclosure of the device's private key.
In the absence of an out-of-band-channel or in the case where the device?s capabilities are  limited, for instance, lack of secure key storage, a secure proxy can be leveraged to handle proper data serialization (\rsec{section:data:serial}).}

\shorten{During this process, neither the owner's nor the device's private keys leave the secure local memory area.}

 \begin{figure}[t]
	\begin{center}
	\includegraphics[width=1\columnwidth]{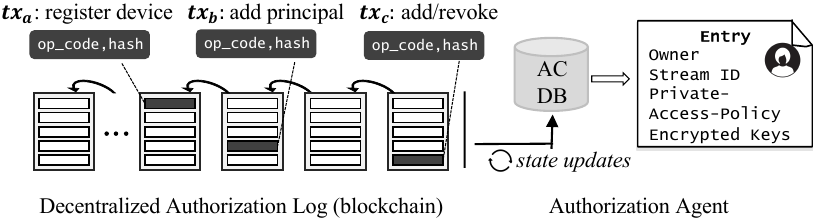}
	\vspace{-15pt}
	\caption{\oursystem's authorization agent bootstraps
	the access control state machine (consolidated into the AC DB) %
	from the \shorten{state} transitions embedded in the decentralized log %
	and accompanied off-chain access policies
	(not depicted for simplification).}
	\vspace{-21pt}
	\lfig{ac-statemachine}	
	\end{center}
\end{figure}

\fakeparagraph{Access Permissions.} 
We utilize the blockchain to store access permissions in a secure, tamperproof, and time-ordered manner.
Access permissions are granted per data stream. %
Initially, the data owner issues a transaction including the stream ID which creates the initial state.
To change this state, e.g., grant read access permissions to a principal, the data owner issues a subsequent transaction which holds, among others,
\textit{(i)}~the stream ID,
\textit{(ii)}~the public key of the principal they want to share their data with,
\textit{(iii)}~the temporal scope of access (e.g., intervals of past or open-end subscription), and
\textit{(iv)}~encrypted keying material for data decryption (\rsec{key:management}).
\blue{For public key discovery of users, \oursystem can leverage decentralized identity management solutions~\cite{onename, Keybase, dif}.
These efforts focus on establishing an open and standards-based decentralized identity ecosystem, removing any reliance on centralized systems of identifiers.
Such solutions, e.g., Keybase~\cite{Keybase}, serve as a key directory that maps a user's online identities (e.g., Twitter, Github) to their public key in a publicly verifiable manner.
The higher the dimensions of interlinked identities, the lower the probability of identity fraud.}

\shorten{\brown{To access data, a principal retrieves the keying material through \oursystem's client and requests the data from the storage.}}
Once a request to store or retrieve data is received at a
storage node, it queries the \oursystem's authorization agent (\rsec{design:virtualchain}) for the corresponding access permissions, as illustrated in \rfig{ac-statemachine}.
To enforce the permissions, the storage node verifies the identity of the requesting user via a signature-based authentication~\cite{chaum1989undeniable}.
Data owners express and dynamically adjust permissions through \oursystem client, which interacts with \oursystem's authorization log only (i.e., blockchain) and not with individual services (i.e., asynchronicity).
Data access is enforced cryptographically through end-to-end encryption.
The storage node validates data access requests based on the embedded access permissions in the authorization log.
The authorization log additionally protects storage nodes' network resources (i.e., bandwidth/memory) from unauthorized users.
For instance, this mitigates an attack, where malicious parties flood the network with download/storage requests of large files.
The storage node can terminate malicious sessions (e.g., data scraping and storage spamming attacks) after checking the access permissions (\rsec{design:virtualchain}).
\blue{\oursystem supports privacy-preserving access permissions  and audits by authorized entities, which we explain in \rsec{sec:dualkey}.}

\fakeparagraph{\blue{Access Policy Indirections.}}
Blockchain storage is scarce and expensive, as it is replicated and maintained by the blockchain network.
This entails placing only the minimum necessary logic in the blockchain.
To keep the number and more importantly the size of transactions as low as possible, our 
design incorporates off-chain storage of the Access Policy, as illustrated in \rfig{transactions}.
Instead of holding the address information of all services, the transaction includes an 
indirection to the Access Policy via the hash digest of it.
This allows managing access permissions with an unlimited number of services in a single 
transaction.
Besides, the Access Policy can now contain advanced access control logic (e.g., 
XACML~\cite{XACML}), such as access groups and delegating parties.
Any change to the Access Policy requires a new transaction.
The hash digest serves as a data pointer and, more importantly, protects the integrity of the Access Policy.
The Access Policy is stored off-chain.
\shorten{The ACL is stored off-chain either in the P2P storage network of \oursystem (\rsec{design:virtualchain}) 
or alternative storage services.}
The time until an access permission change comes into effect is tied to the transaction confirmation time of the underlying blockchain, ranging from few seconds to minutes depending on the underlying blockchain.

 \begin{figure}[t]
	\begin{center}
	\includegraphics[width=0.8\columnwidth]{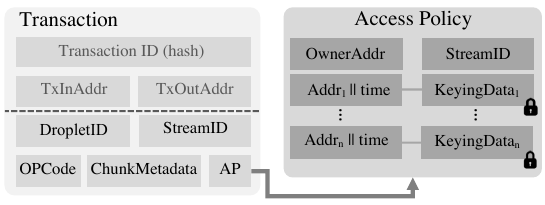}
	\vspace{-10pt}
	\caption{Overview of access control transactions, which
	 embed transitions to the global access control state via an indirection (i.e., hash to the off-chain Access Policy).
	}
	\vspace{-15pt}
	\lfig{transactions}	
	\end{center}
\end{figure}

\vspace{-5pt}
\subsection{\blue{Privacy-Preserving Sharing}}
\lsec{sec:dualkey}
\vspace{-5pt}
In public blockchains, users are represented through virtual addresses, providing pseudonymity.
However, advanced clustering heuristics can potentially lead to the de-anonymization of users~\cite{meiklejohn2013fistful, androulaki2013evaluating}.
Access permissions in \oursystem should be enforceable by storage nodes (i.e., verify authorization) and be auditable by authorized parties.
However, we want to protect the privacy of sharing relationships from the public.
To realize this, we leverage dual-key stealth addresses.
\blue{With stealth addresses~\cite{stealth}, we protect a principal's privacy, from any party who can view the access permissions, with regards to the resources they are granted access to.}
Moreover, different streams shared with the same principal are unlinkable.
\blue{However, a data owner may learn whom they are sharing their data with.}
\blue{Note that if there is no channel between the data owner and data consumer to indicate requested or granted access permissions, 
the consumer needs inevitably to scan the permissions in \oursystem's access permission state machine to identify any data that is shared with them.}

Conceptually, each principal is represented by two public keys (main and viewer keys: $PK_{m}$, $SK_{m}$, $PK_{v}$, $SK_{v}$),
which other parties use to generate a new unlinkable address $PK_{new}$.
The viewer private key $SK_{v}$ can be shared with an auditor to audit the permissions.
However, access to both main and viewer private keys is required for data access,
i.e., $SK_{m}$ and $SK_{v}$ are needed to compute $SK_{new}$, which only the principal is capable of  (see \rap{app:stealth} for technical details).

\vspace{-5pt}
\subsection{Access Control State Machine}
\lsec{design:virtualchain}
\vspace{-5pt}

Today, there are two main options developers can take for realizing decentralized applications that employ a blockchain as a ubiquitous trust network (i.e., a shared ground truth):
\textit{(i)}~operating a new blockchain, or
\textit{(ii)}~embedding the application logic into an existing secure blockchain deployment~\cite{catena, Virtualchain}.
We opt for the latter where we embed our logic without alternation of the underlying blockchain \blue{nor requiring the instantiation of a new blockchain}. 
This allows us to benefit from an existing blockchain's security properties without introducing and running a new blockchain. \shorten{and make our design generic.}
\blue{Note that \oursystem's state machine can alternatively employ a private authorization log, to address use-cases with a different trust model or in a closed ecosystem.}
We briefly discuss the reasons why we opt for this choice and detail on how we realize this efficiently.

Integrating a new application logic into a running blockchain typically results in consensus-breaking changes and hard forks\shorten{~\cite{Blockstack}}, i.e., a new blockchain with only a subset of peers enforcing the new logic. %
While necessary for specific applications, this results in parallel blockchains which may not exhibit strong security properties due to a smaller network of peers
\blue{(e.g., Namecoin's network became decentralized with one mining group controlling the majority of hashing power~\cite{Blockstack}).}
To benefit from the security properties of a strong and robust blockchain,
new apps can embed their log of state changes in transactions\shorten{~\cite{Blockstack}}. %
This is in turn used to bootstrap the global state in a secure and decentralized manner.

We employ \shorten{the approach of} virtualchains~\cite{Virtualchain, Blockstack} to efficiently embed \oursystem's access control logic  in \blue{an existing global blockchain.}
A virtualchain is a fork*-consistent 
replicated state machine, allowing different applications to run on top of any production blockchain, without breaking the consensus.
\oursystem's authorization agent scans the underlying blockchain for the corresponding access permission transactions and maintains the global state in a database
that can be queried for permissions of a given stream and principal.

 \begin{figure}[t]
	\begin{center}
	\includegraphics[width=0.9\columnwidth]{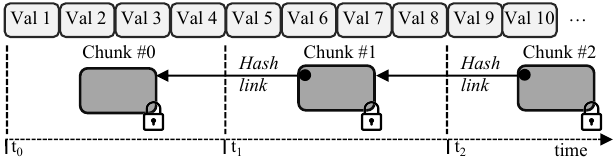}
	\vspace{-14pt}
	\caption{Data streams chunked at defined temporal intervals, and cryptographically linked together.
	For record lookup, the timestamp is mapped to the chunk identifier.}
	\vspace{-21pt}	
	\lfig{datachunking}	
	\end{center}
\end{figure}

\vspace{-5pt}
\section{\blue{Data Serialization}}
\lsec{section:data:serial}
\vspace{-5pt}
\lsec{data:serialization}
In \oursystem's data model, a data stream is divided into chunks of predefined time intervals (\rfig{datachunking});
chunking and batching are common techniques for time-series data~\cite{Bolt,time-series-Freedman, chronix, minicrypt}.
\shorten{Instead of storing individual data records, we store data chunks, which are an ordered batch of data records of an arbitrary type (i.e., pairs of timestamp/value).}
Although chunking prevents random access at the record level, 
it results in a positive performance gain for data retrieval as in time-series data most queries require access to temporally 
co-located data~\cite{Bolt, minicrypt}.
E.g., data analytic apps work with temporal data records (e.g., all records of a day).

\fakeparagraph{Encryption.}
Each data chunk is initially compressed and then encrypted at the source with an efficient symmetric cipher\footnote{
Note that it is important to apply padding to prevent inference attacks based on the varying sizes of the chunks.}.
We rely on AES-GCM, as an authenticated encryption scheme.
Note that NIST bounds the use of AES-GCM to $2^{32}$ encryptions for a given key/nonce pair.
Due to our frequent key rotations, we stay far below this threshold.
The chunks %
have a metadata segment containing, among others, the chunk identifier, the owner's address,
hashes to previous chunks (\rsec{design:immut}),
and the stream identifier.
The data field contains the encrypted and compressed data records.
Services with access to the encryption key can verify the integrity of the chunk and perform an authenticated decryption.
To ensure data ownership,\shorten{\red{for instance to the storage node},} each chunk is also digitally signed.
This allows parties without access to the encryption key \shorten{to still be able}
to verify the owner of the data stream, albeit at a higher computation cost. 
\shorten{In general, digital signature operations are three orders of magnitude slower than symmetric key operations, as discussed in~\rsec{eval:data}.}

\fakeparagraph{Storage Interface.}
The storage nodes\shorten{in \oursystem} expose a key-value interface, with a common \textit{store}/\textit{get} interface with various flavors of \textit{get}, such as \textit{getAll} or \textit{getRange}.
For each incoming request, the storage node first verifies the identity of the client (i.e., authentication) and looks up the corresponding access permissions regarding the client's identity (i.e., authorization).
Each request is accompanied with a universally unique identifier (UUID), defined as the hash of the tuple: $<$\texttt{owner 
address, streamID, counter}$>$, where \texttt{streamID} is a unique identifier of an owner's data stream.
Traditional indexing for data retrieval cannot be applied here as data chunks are encrypted.
Hence, we need to devise a mechanism to perform temporal range queries over encrypted data efficiently.
To avoid consistency issues of a shared index, we exploit a simple local lookup mechanism to enable temporal range queries.
For a constant lookup time of a record with timestamp $t_i$, we compute the counter of the chunk holding it based on the known time interval $\Delta$ of the chunks:
$\lfloor (t_i-t_0)/\Delta \rfloor$.
For instance, we can map the lookup of value 7 in \rfig{datachunking} to the identifier of chunk $\#$1.
The chunk metadata is included in the initial stream registering transaction, as depicted in \rfig{transactions}.
Note that the chunk metadata additionally enables freshness checks for chunks, since
the chunk interval indicates the frequency and time at which new data chunks are generated.

\fakeparagraph{Strong Data Immutability.}
\lsec{design:immut}
While \oursystem provides integrity protection via authenticated encryption and digital signatures, the data owner can still modify old data.
Specific applications might require a stronger notion of immutability such that even the data owner can no longer modify the data (e.g., contractual agreements in logistics).
\oursystem enables such a notion of immutability through blockchain's append-only property~\cite{BitCoin:SoK}.
The application developer can define a grace period, after which data chunks become immutable. 
For sensitive applications, this can be per chunk.
Otherwise, a more extended period can be selected.
To accommodate for the narrow bandwidth of blockchains, we leverage an anchoring technique, where data immutability transactions are reduced to the level of the grace period.
To realize this, the first data chunk holds a pointer to the registration transaction, and after the grace period, a transaction with a pointer to the latest chunk is issued, as depicted in~\rfig{datachunking:immut}.
Since all data chunks are cryptographically linked via hashes, all data chunks in the grace period become immutable at once, forming a chain of data chunks.
To avoid a linear verification time, chunks hold hashes to several previous chunks, forming a geometric series (i.e., logarithmic verification time).

\vspace{-10pt}
\section{\blue{Privacy and Security Analysis}}
\vspace{-10pt}
\lsec{design:privacy}
\fakeparagraph{Authorization.}
In conventional authorization frameworks, i.e., OAuth,
any entity in possession of the bearer token can assume the same access permissions granted 
to the token~\cite{sun2012devil}.
In case of token theft, the adversary in possession of the token can gain unauthorized access 
to the user's resources (i.e., impersonation attack).
Moreover, the compromise of an authorization server enables the issuance of unauthorized 
access tokens for all registered resources at the authorization server.
\oursystem is not susceptible to these attacks.
In \oursystem,
an authorization claim with the scope of access is logged in the blockchain in a privacy-
preserving manner, such that only the authorized party in possession of the correct private key 
can claim ownership for data access, in a publicly-verifiable manner.
For an adversary to alter access permissions in the blockchain, it requires forging a digital 
signature (i.e., breaking public key 
cryptography with a 128-bit security level) or gaining control over the majority of the computing 
power in the blockchain network (i.e., 51\% attack~\cite{Blockstack}).
Existing production blockchains, e.g., Bitcoin or Ethereum, can be subject to security attacks, 
such as 
routing~\cite{hijacking} and selfish mining~\cite{eyal2014majority}, which can lead to access 
permission state update transactions to be dropped, delayed, or excluded. 
An active adversary can employ these attacks to prevent/delay access permission 
modifications of victims from taking effect. 
However, none of these attacks can lead to unauthorized access permission.

 \begin{figure}[t]
	\begin{center}
	\includegraphics[width=1\columnwidth]{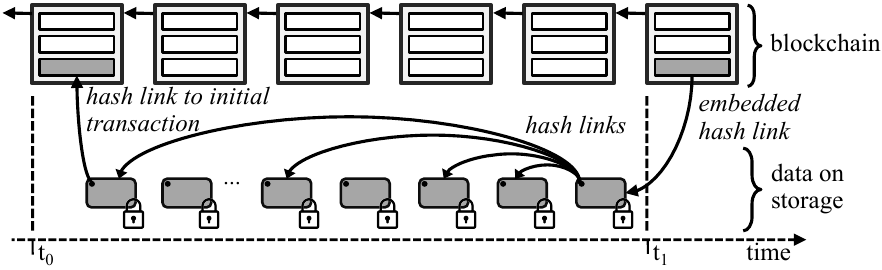}
	\vspace{-25pt}
	\caption{Example of immutable chunks, with a grace period ($t_1 - t_0$).
	Chunks are cryptographically linked together, forming a geometric series, enabling faster integrity checks.}
	\vspace{-22pt}	
	\lfig{datachunking:immut}	
	\end{center}
\end{figure}

An adversary is not capable of learning sensitive information from the public blockchain,
since only unlinkable pseudo-identities and stream identifiers are stored. 
In profiling attacks, the adversary creates profiles of all user identifiers and the network of users~\cite{meiklejohn2013fistful}.
An adversary can break the pseudonymity of specific users. %
Hence, a large body of research aims at concealing identity and relationships in public blockchains while maintaining 
verifiability~\cite{sasson2014zerocash, bulletproofs, green2017bolt}.
\oursystem employs dual-key stealth addresses, where the anonymity set is equal to the set of users using 
non-spendable stealth addresses.

A malicious storage node \blue{(or authorization agent)} could hand out data without permission or data leakage might take place due to system compromise.
However, the impact of this action is limited since data is end-to-end encrypted.
Moreover, leakage of a data encryption key results only in the disclosure of the data stream segment associated with it.
The compromised key cannot be used to disclose old data nor can it be used to gain access to future data due to pre-image resistance property of hash functions.
The distribution key (KD) for continuous stream subscription gives access to the latest token from the primary chain.
The compromise of KD has no impact without access to the aligned token from the secondary chain (\rfig{keyregression}) since both tokens are required to compute data encryption keys.
An attacker needs to compromise an authorized user's private key to gain access to tokens from the secondary chain.
The blockchain provides auditable information about when a stream was shared with whom; 
a crucial piece of evidence to prove/disprove access rights violations should the need arise.

\fakeparagraph{Data Serialization.}
Data chunks are encrypted, integrity protected, and authenticated.
Any data chunk manipulations are detectable via the digital signature and authenticated encryption.
Note that while a property of AES-GCM can be exploited to find collisions within ciphertexts that decrypt to different valid plaintexts~\cite{dodis2018fast},
the per chunk signature in \oursystem protects us from such an attack.
The optional data immutability is based on the security and immutability of blockchain.
The secure channel (i.e., TLS) for storing and fetching data prevents replay attacks, 
in addition to ensuring an authenticated and confidential channel.
An adversary with access to disclosed encryption keys cannot alter old data, as it requires access to the signing private key.

\vspace{-5pt}
\section{Implementation}
\vspace{-5pt}
\lsec{implementation}

Our reference implementation of \oursystem is composed of three entities implemented in Python: the client engine, the storage-node engine, and the authorization agent.
The client engine is implemented in 1700~sloc.
We utilize Pythons's \texttt{cryptography} library~\cite{pcl} for our crypto functions. %
For compression, we use \texttt{Lepton}~\cite{dropboxlep} for images and \texttt{zlib}~\cite{zlip} for all other value types.
\begin{figure*}[t!]
 \centering
       \subfigure[Average throughput for \textit{get}.]{   
    	\lfig{eval:throughput}
    	\includegraphics[width=0.98\columnwidth]{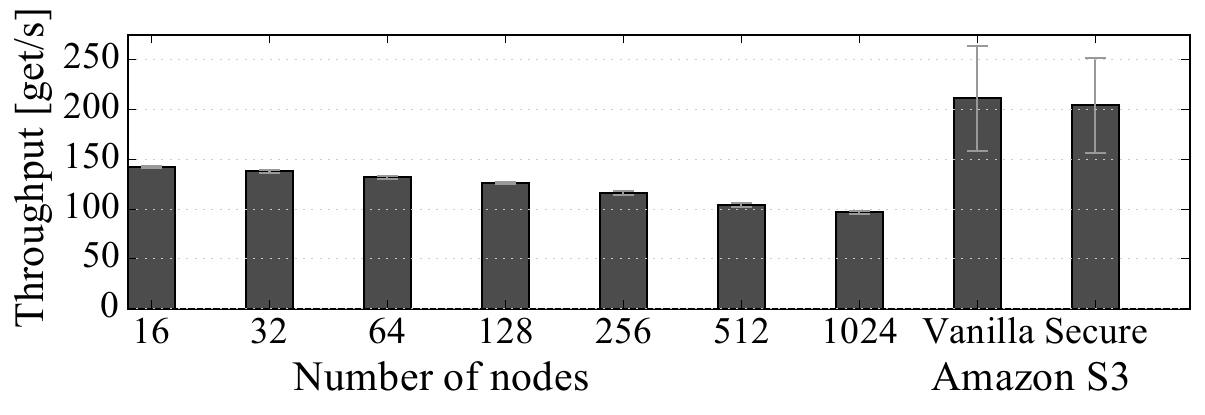}
       }
 \subfigure[Latency for single \textit{store} and \textit{get} requests.]{ 
    \lfig{eval:latency}
    \includegraphics[width=0.98\columnwidth]{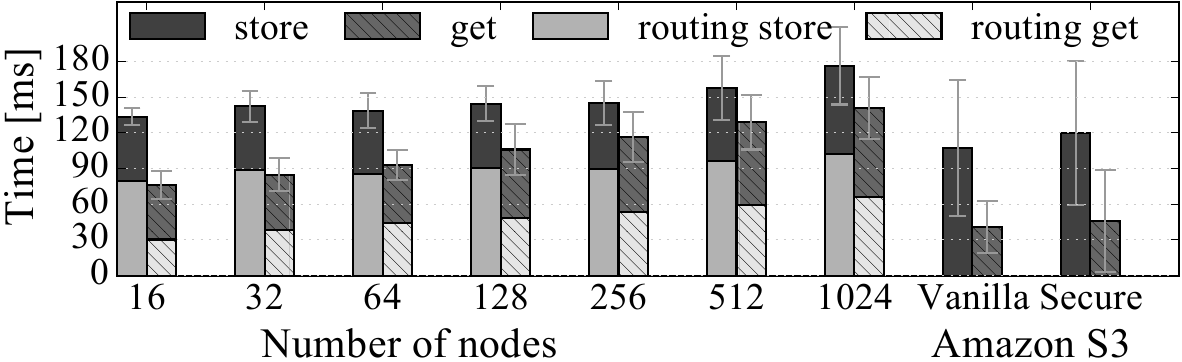}
    } 
	\vspace{-12pt}
    \caption{\textit{store}/\textit{get} performance for centralized and decentralized storage layers.
	The latency for the decentralized storage is dominated by network routing.
	For fairness, all settings, including Vanilla S3 (w/o \oursystem) operate on compressed data chunks.
	\lfig{benc}
	}
	\vspace{-10pt}
\end{figure*}

The storage engine can either run on the cloud or nodes of a p2p storage network.
For the cloud, we have integrated a driver for Amazon's S3 storage service.

We have as well a realization of \oursystem with a serverless computing platform 
with ASW Lambda serving as the interface to the storage (i.e., S3).
Once Lambda is invoked, it performs a lookup in the access control state machine to process the authorization request.
For comparison, we implement as well an OAuth2 authorization, based on AWS Cognito~\cite{cognito}.
For the distributed storage, we build a DHT-based storage network.
We instantiate a Kademlia library~\cite{pythdht} and %
extend it with the security features of S/Kademlia~\cite{SKademlia}.
On the p2p storage nodes, we employ LevelDB~\cite{levelDB}.
Our extensions amount to 2400~sloc.

The authorization agent is implemented with the virtualchain library~\cite{Blockstack} to maintain the access control state 
machine.
The virtualchain scans the blockchain, %
filters relevant transactions, validates the encoded operations, %
and applies the outcome to the global state.
The state is persisted in an SQLite database.
The global state can either be queried through a REST API or accessed directly through the SQLite database.
Our extensions to the virtualchain amount to 1400~sloc.
As the underlying blockchain, we employ a Bitcoin test-network with a block generation time \shorten{to emulate a hybrid consensus blockchain~\cite{OmniLedger}} of 15~s.%

\begin{table}[tb]
  \begin{center}
  \footnotesize
   \begin{tabular}[b]{lcccccc}
		\toprule
& \multicolumn{2}{c}{AES Encrypt} & \multicolumn{2}{c}{SHA Hash} & \multicolumn{2}{c}{ECDSA Sign} \\
		\cmidrule(lr){2-3}
		\cmidrule(lr){4-5}
		\cmidrule(l){6-7}
&						 [$\mu$s] & [op/s]  & [$\mu$s] & [op/s] &  [ms] & [op/s]\\
	    \midrule
 IoT SW&            	 298 & 3.4k & 297 & 3.4k & 270 & 3.7\\ %
 IoT HW &            	 42 & 23.8k & 17 & 58.8k & 174 & 5.7\\ %
 Phone	&          50 & 20k & 45 & 22.2k & 4.4 & 227\\
 Laptop	&          5.4 & 185k & 1.6 & 623k & 1.3 & 770\\
 Cloud &          2.6 & 384k & 1.2 & 833k & 1.1 & 909\\
		\bottomrule
        \end{tabular}
  \end{center}
  \vspace{-15pt}
  \caption{Performance of security operations -- 128-bit security. 
  For IoT devices, we use OpenMote microcontrollers with software~(SW) computations or crypto\shorten{~hardware} accelerators~(HW). 
  As a smartphone, we use a Nexus 5.
  As a laptop, we use Macbook Pro. 
  For the cloud, we use an Amazon t2.micro instance.
  }
  \ltab{eval:crypto}
  \vspace{-1pt}
\end{table}

\vspace{-5pt}
\section{Evaluation }
\lsec{eval}
\vspace{-5pt}

\fakeparagraph{Goals.} 
One of our primary goals was to develop \oursystem as a practical system,
which translates to ensuring: that \oursystem can  
\textit{(i)}~be supported by existing resource-constrained IoT devices, 
\textit{(ii)}~sustain a high access permission lookup and verification throughput, and 
\textit{(iii)}~that the overhead to both data owners and consumers is low, allowing consumers to process large volumes of data streams. 
Hence, our evaluation metrics include the overheads (CPU, memory) that \oursystem imposes on each party, as well as the end-to-end throughput and latency that apps experience with \oursystem.
Our evaluation is conducted in the context of real-world devices, datasets, and runtime environments.

\fakeparagraph{Devices.}
We perform our evaluation on the following four device classes:
\textit{(i)}~IoT: OpenMotes equipped with 32-bit ARM Cortex-M3 SoC at 32~MHz, 
a public-key crypto accelerator running up to 250~MHz.
Fitbit trackers utilize a similar class of micro-controllers;
\textit{(ii)}~smartphone:  LG Nexus5 equipped with a 2.3~GHz quad-core 64-bit CPU, 2~GiB RAM; 
\textit{(iii)}~laptop: MacBook Pro equipped with 2.2~GHz Intel i7, 8 GiB RAM;
\textit{(iv)}~Cloud: EC2 t2.micro (1~vCPU, 1~GiB RAM).

\fakeparagraph{Datasets.}
We validate the applicability of \oursystem by deploying three real-world IoT applications atop of \oursystem and
quantifying the end-to-end overhead due to our system;
\textit{(i)}~for the Fitbit activity tracker, we use the anonymized fitness tracker data of the co-authors over one year (16 data types, 130~MB), which we use to synthesize data for an arbitrary number of users.
\textit{(ii)}~for the Ava health tracker~\cite{avawomen}, we use an anonymized dataset from Ava~\cite{avawomen} (10~s intervals, 13 sensors, 1.3~GB). 
\textit{(iii)}~for the ECOviz smart meter dashboard, we use the publicly available anonymized ECO dataset (1.85~GB) for 6 households over 8~months~\cite{beckel2014nilm}.

\fakeparagraph{Storage Infrastructure and Runtime Environment.}
We run \oursystem on both centralized and decentralized storage layers.
For the former, we use Amazon's S3 service, and for the latter, we implement 
and run several DHT nodes in real-time on an emulated network 
(e.g., using \texttt{netem}~\cite{netem}). 
Evaluating \oursystem in a decentralized storage setting is a compelling case, as peer-to-peer storage networks could become a viable solution for the IoT~\cite{GDP}.
Additionally, this setup resembles storage-oriented blockchains (e.g., Storj~\cite{storj}, Filecoin~\cite{filecoin}),
which still lack adequate mechanisms for secure data sharing, where \oursystem can be helpful.
We also evaluate \oursystem's performance in a serverless setting (Lambda~\cite{lambda}) and compare it to OAuth2 authorization. %
Emerging serverless platforms require request-level authorization~\cite{adzic2017serverless}, 
where \oursystem can serve as an \textit{Authorization as a Service}.

\vspace{-5pt}
\subsection{Microbenchmark}
\lsec{eval:micro}
\vspace{-5pt}
We instrument the client engine to perform the microbenchmark in isolation with up to 1000 repetitions.

\lsec{eval:data}

\fakeparagraph{Cryptographic Operations.}
\rtab{eval:crypto} summarizes the costs of the crypto operations involved in \oursystem on four different platforms. %
All these operations, namely AES encryption, SHA hash, and ECDSA signature are performed once per chunk for store requests.
For data retrieval, the client does not perform a signature verification, since AES-GCM has built-in authentication.
Running the crypto operations only in software on the IoT devices shows the highest cost, with 3.4k~encryptions/hashes 
per sec and only 3.7~signatures per sec.
With the onboard hardware crypto, the cost of AES and SHA is improved by one order of magnitude and approaches that
of smartphones.
Note that overall signatures are three orders of magnitude slower than symmetric key operations.

\fakeparagraph{Crypto-based Access.}
Hash computations are the basis for dual-key regression.
The computation occurs at the initial setup and each key update if the client chooses to re-compute keys on-demand rather than store them.
Assuming a chain length of 9000 (hourly key updates for one year), it takes 405~ms to compute the entire chain on smartphones and 2.7~s on an IoT device without a hardware crypto engine. %
With compact hash chains, we reduce this worst-case compute time to 4.3 and 28.2~ms, respectively.
The performance gains become pronounced with smaller epoch intervals.
The hash tree induces $O(log~n)$ computations for $n$ keys, which amounts to 48~$\mu$s (laptop) with $2^{30}$ keys. %

The per chunk overhead consists of key computation (hash tree and dual-key regression), chunk encryption, key encryption, and signature, which amounts to 1.5~ms (laptop) without caching.
Compared to ABE (\rsec{relatedwork}), \oursystem's crypto-based data access is by a factor of 57x faster.
E.g., with ABE per chunk overhead with only two attributes (timestamp for temporal access and data type) amounts to 86~ms (laptop).

\fakeparagraph{\blue{Feasibility for IoT.}}
To assess if Droplet is viable for the IoT, we validate its practicality for low-power devices, concerning their constraint resources (\rtab{eval:crypto}). 
Crypto operations are the most expensive ones on a data producer, and beyond that, no connectivity to the authorization services is required.
Today, most IoT devices are equipped with crypto accelerators for AES encryption integrated with their radios;
however, accelerators for hash functions and signatures have yet to\linebreak become the norm.
Nevertheless, Droplet is feasible on legacy IoT devices without accelerators despited 1.5x slower signing operations.
In terms of impact on the energy budget, the signature consumes only~9 to 25mJ.
Considering a wearable's lithium-polymer battery capacity of 1.2~Wh (4.32~kJ), and a 48h charge cycle, 3  signatures/minute (8.6 with accelerator) can be computed with 5\% of the energy budget.

\vspace{-5pt}
\subsection{System Performance}
\lsec{eval:system}
\vspace{-5pt}
To model the real-world performance of \oursystem, we constructed an end-to-end system setup, where we use our three apps datasets.
Note that we do not cache any data to emulate worst-case scenarios. %
The stream chunk size is set to 8~KiB.
We evaluate \texttt{get} and \texttt{store} requests to the storage layer, which include the overhead of \oursystem's access control. 

\fakeparagraph{Serverless Computing.}
In the serverless setting, Lambda either runs \oursystem for the access control or uses the AWS Cognito service, which runs OAuth2, as the baseline.
Lambda with both Droplet and Cognito exhibits a latency of around 118~ms (0.4\% longer with Droplet).
Note that with OAuth2, to reach the same level of access granularity as with \oursystem, separate access tokens are required for each data chunk, which is impractical.
This is why in practice, long-lived and more broadly-scoped access tokens are granted.

\fakeparagraph{Cloud.}
We extend AWS S3 storage with \oursystem and compare its performance against vanilla S3. %
\rfig{eval:throughput} shows the throughput for different request types.
We follow Amazon's guidelines to maximize throughput: e.g., the chunk names are inherently well distributed allowing the best performance of the underlying hash-table lookup.
The vanilla S3 throughput of 211~gets/s is within Amazon's optimal range (100-300).
With \oursystem, we maintain an average rate of 204~get/s (3\% drop).
\rfig{eval:latency} shows the latency for individual \textit{store} and \textit{get} operations.
In \oursystem, the latency overhead is 13\% for \textit{get} and 11\% for \textit{store} (incl. crypto). %
Part of the overhead is due to the expensive signature operation. %
Also, there is an overhead for a fresh lookup of access permissions at the access control DB of the virtualchain\shorten{~instance}. %

\fakeparagraph{Distributed Storage}
We measure the performance of \textit{get} and \textit{store} requests on a secure DHT with \oursystem, with varying network sizes, from 16 to 1024 nodes.
\rfig{eval:throughput} shows the throughput results.
As the number of nodes increases from 16 to~1024, the performance decreases from 142 to 96~get/s.
\rfig{eval:latency} shows the latency results, divided into routing and retrieval.
The total \textit{get} latency increases from 76 to 140~ms as the number of nodes grows.
This is about 3 times slower than S3's centralized storage.
However, note that the routing cost dominates this slowdown.
After resolving the address of the storage node, which holds the data chunk, the secure retrieval time is similar to that of S3.
Also, note that \textit{get} requests have a lower routing overhead than store requests.
This is because for get requests, the routing process is aborted as soon as a node holding the data chunk is found.

\shorten{
\fakeparagraph{Off-Chain Metadata Cost Estimations.}
Access policies are unique per stream and per owner. 
Considering a user-base of one billion, each with an average of 100 streams, we end up with 10T access permissions.
With an average of 100 entries per access permission
(i.e., 10~KB), the total size of storage required for the access permissions accumulates to 100~PB.
This data must be persisted on a reliable and available storage layer.
Considering today's pricing of AWS S3 storage\footnote{S3 frequent access tier, over 500~TB/Month, \$0.021/GB, May 2020.}
the average annual cost per 4000 users is at 1\$.
}

 \begin{figure}[t]
	\begin{center}
	\includegraphics[width=1\columnwidth]{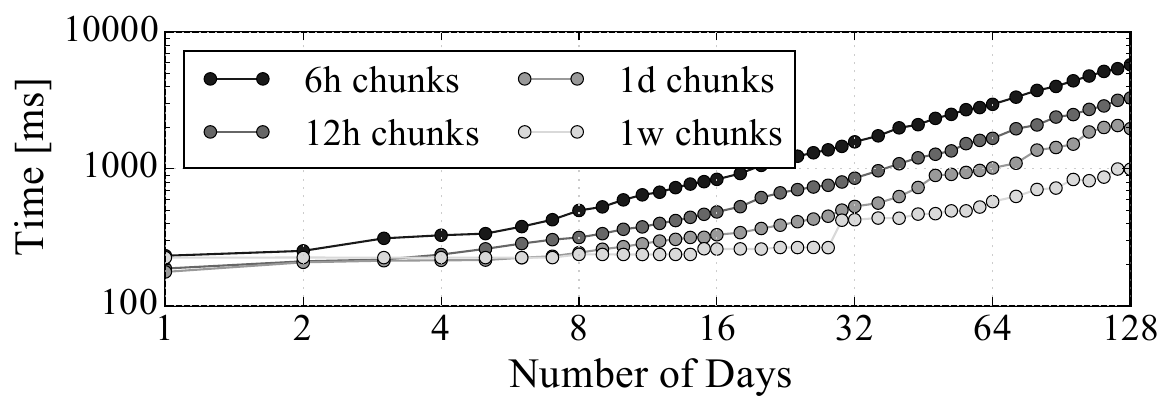}
	\vspace{-22pt}
	\caption{EccoViz app results.
	Retrieving records from the energy data set in the EccoViz dashboard app (p2p storage).}
	\vspace{-22pt}
	\lfig{systembench:apps}	
	\end{center}
\end{figure}

\fakeparagraph{Applications.}
The three applications we deploy atop \oursystem, vary in terms of type, size, and granularity of collected data.
Fitbit and Ava are both smartphone apps, where users view visualized summarizes about their collected data and set goals.
We enhance both demo apps, with additional views where the user can selectively share parts of their data (e.g., heart-rate/body-temperature/steps) with friends or services over \oursystem.
ECOViz dashboard is a web app that visualizes energy consumption from smart-meters.
Users can set access permissions per data stream, and they can only view streams to which they have been granted access.
The user experience of sharing via \oursystem remains similar to that of existing sharing methods.
Users initially register a data stream either consisting of a single or multiple data types (e.g., sensitive data types can be highlighted to prevent accidental sharing).
Afterward, they can add or remove users to/from their data streams 
(e.g., the iOS native Health app allows per data stream sharing decisions for third-party apps, similar to our subscription mode).

To measure the overhead induced by \oursystem, we quantify the overhead of \textit{store} and \textit{get} data requests for different views (i.e., each access requires cryptographic operations and access permission checks).
We now discuss the decentralized storage setting with 1024~nodes.
Due to memory constraints, data synchronization is required at least weekly for Fitbit and daily for Ava devices.
This results in an average \textit{store} latency of 176~ms and 1.2~s for Fitbit and Ava, respectively.
Note that \textit{store} operations run in the background.
For different views, the maximum \textit{get} latency is below 150~ms.
Hence, the user experience remains unaltered. 

In contrast to Fitbit and Ava, the smart meter node has direct Internet connectivity.
Instead of synchronizing periodically, it stores chunks after generation. %
This takes 176~ms per chunk.
The most comprehensive view in the ECOViz dashboard can visualize the entire data stream.
\rfig{systembench:apps} shows the latency to fetch chunks dependent on the number of days requested.
Fetching data for 128~days of 6~h chunk size takes about 10~s, whereas the one-week size takes less than 1~s.

\fakeparagraph{Scalability.}
\oursystem's scalability can be examined from three angles;
\textit{(i)~Read throughput of authorization};
read operations are performed in $O(1)$, after the authorization agent bootstraps the Access Control State Machine DB.
Scaling to handle high read throughputs, is a matter of increasing the number of authorization agents.
\textit{(ii)~Storage of access permissions};
\oursystem anchors indirections in the blockchain (\rsec{design:blockchain}), as we store\linebreak access policies and metadata off-chain.
Hence, to scale with the growing number of access permissions, the allocated off-chain storage is dynamically increased.
As \oursystem scales to a more significant number of data streams, the access permission logic consequently grows.
The individual authorization agents are not impacted by this growth,
as they only store the state for the resources they serve.
\blue{The annual meta-data storage costs\footnote{\blue{S3 frequent access tier, over 500~TB/Month, \$0.021/GB, May 2020.}} 
for a billion user-base with an average of 100 streams and 100 consumers per stream, would amount to less than \$0.001 per user today, which accounts for a fraction of the actual storage costs of streams.} %
\textit{(iii)~Write throughput};
represents the scaling bottleneck of \oursystem, as access permission updates are bound to the write throughput of the underlying blockchain.
Although we consider several optimizations (e.g., grouping access updates) to contain this constraint, it remains a bottleneck.
In our current prototype, the transaction confirmation time is set to 15~s, similar to that of Ethereum.
The slow blockchain writes have a direct impact on the time until new access permissions take effect, which is significantly higher compared to OAuth2 protocol.
Read-throughput is, however, fast and comparable to that of OAuth2.
Data stream registrations and access permission adjustments (e.g., grant/revoke access) require transaction writes.
To understand the extent of scaling authorization writes in \oursystem with an example, consider Fitbit with 25 Million active users, which logged 4.7 million group-join events in 2017~\cite{fitbitdata}, which would require 0.14~transactions per (tps).
However, to scale \oursystem to billions of data streams, a blockchain throughput of a few thousand tps is necessary (assuming 25\% of streams require an access permission modification per day).
While currently deployed blockchains achieve only a fraction of this throughput, scaling to higher throughput is an active area of research, and next-generation blockchains already support several thousand tps~\cite{OmniLedger}(\rsec{discussion}).

\vspace{-5pt}
\section{Discussion} %
\lsec{discussion}
\vspace{-5pt}

We highlight some research questions that remain open.

\fakeparagraph{Beyond IoT.}
An authorization service with \oursystem's properties is crucial for systems that advocate for data sovereignty~\cite{GDPR-LAW,guardian-thielman,GDP}
or handle privacy-sensitive data,
e.g., sharing medical records~\cite{Medrec}, and humanitarian aid~\cite{Digital-Immunity}. 
The storm of recent privacy  incidents~\cite{fb-scandal, techcrunch-equifax} has prompted a rethinking of this space.
Moreover, decentralized storage services that run on blockchain (e.g., Filecoin) can integrate \oursystem for data sharing.
Services with varying trust assumptions can, however, run \oursystem's authorization log instead by a federated set of servers. \shorten{(i.e., permissioned blockchain).}

\shorten{
With \oursystem, we bring an alternative design that is crafted to give users full control and sovereignty over their data, without relying on trusted intermediates. 
This might be an extreme departure of the norm today, however, we hope \oursystem brings interesting insights and discussions 
to an important challenge; namely authorization and fine-grained access control for end-to-end encrypted  data streams.
}

\shorten{
\fakeparagraph{Beyond Streams.} 
This paper addresses the challenge of flexible and 
granular authorization, sharing, and 
accessing of stream data in an end-to-end encrypted setting without the need for centralized trusted entities to facilitate that.
The decentralized authorization service and crypto-enforced access are  two key components designed to achieve this particular goal.
Our authorization is not tight to stream data and could be coupled with other crypto access control schemes (e.g., ABE) for other types of data, e.g., large files.
}

\fakeparagraph{Usability.}
\oursystem is a user-centric system that empowers data owners with control over their data.
While we design \oursystem's API to abstract away system complexities from users and mimic current data sharing abstractions,
some usability considerations remain open in this user-centric paradigm.
In this paradigm, users will potentially be confronted with more decisions to make regarding their data.
Hence, it is essential to study and design abstractions and interfaces that mitigate usability concerns that might arise in this paradigm.
\shorten{In this design paradigm, the data owner will have to make and manage granular decisions regarding their data. 
We acknowledge that this is a challenge that needs to be addressed with practical abstractions.}
In an end-to-end encryption model, protection and recovery mechanisms for private master keys should be 
addressed with adequate solutions.
For instance, Shamir's secret sharing scheme~\cite{shamir1979share} allows reconstruction of the secret from a set of recovery 
keys which 
are, e.g., distributed among the data owner's devices~\cite{Sieve} or a group of friends~\cite{Preveil}. 
The recovery keys collectively reconstruct a master secret key.

\fakeparagraph{Blockchain Scalability.}
In \rsec{eval}, we discussed scalability aspects of \oursystem and how the underlying blockchain, which realizes the decentralized authorization log, can impact the write throughput within \oursystem. 
Next-generation blockchains~\cite{gilad2017algorand, buterin2017casper, bitcoin-ng, miller2016honey, OmniLedger, kogias2016enhancing} particularly tackle the scalability aspects and
promise higher throughputs and lower latencies, which is crucial for the adoption of blockchain-based systems in retail payments and financial sector, and for realizing large-scale decentralized applications.
Recent works~\cite{kogias2016enhancing, OmniLedger}
introduce a hybrid consensus by combining the slow 
PoW to bootstrap the faster PBFT algorithm,
where for each epoch, a random set of validators is selected.
Hence, they bring both worlds' best: secure open enrollment and high throughput and low latency.
These scalable blockchain protocols, e.g., %
OmniLedger~\cite{OmniLedger}, lay the groundwork enabling practical advanced decentralized services, such as \oursystem.
\oursystem can be deployed on top of any blockchain that supports the total ordering of transactions, as elaborated in~\rsec{design:virtualchain}.

\vspace{-5pt}
\section{Related Work}
\lsec{relatedwork}
\vspace{-5pt}

We now briefly discuss key relevant works to \oursystem.
\fakeparagraph{Crypto-enforced Data Access.} 

End-to-end encryption provides the strongest
level of protection for data stored in the cloud, as data remains encrypted and only authorized entities are trusted with decryption keys.
However, fine-grained access and sharing of data is a challenge here.
A simple approach to selective sharing of encrypted data is to encrypt the target data towards the principal's 
public key;  although simple this approach suffers from three drawbacks:
\textit{(i)}~hard-coded access control~\cite{Digital-Immunity};
at encryption time the access permission is defined and cannot subsequently be altered or revoked,
\textit{(ii)}~storage overhead; if the same data is shared with multiple principals, the user ends up storing redundant data as 
she needs to encrypt the same data under each principal's public key, and 
\textit{(iii)}~scalability and practicality issues particularly when considering fine-grained access policies.
These drawbacks are pronounced with time series-data, where high volume of data is continuously produced and a high key-rotation is necessary to ensure flexible access control. %

Various cryptographic schemes~\cite{boldyreva2008identity, Ateniese:NDSS} have been introduced to overcome some of these challenges, 
among which
attribute-based encryption (ABE)~\cite{goyal2006attribute, Sieve, abe1, abe3, agrawal2017fame} offers high expressiveness. %
\shorten{ABE encrypts data towards a policy (i.e., associated with a set of attributes),
and only principals with the secret key satisfying the policy can decrypt the data.}
Several ABE-based systems~\cite{Sieve, yu2010achieving} introduce crypto-based access control. %
However, ABE suffers from expensive crypto operations %
and the costs grow linearly with the number of attributes, limiting the granularity of access due to computational burdens~\cite{abe5,agrawal2017fame}.
The overhead dominates even with a hybrid encryption technique~\cite{Sieve, yu2010achieving}, where %
data is encrypted with symmetric encryption and only encryption keys are encrypted with the expensive ABE,
e.g., only two attributes result in 100~ms for enc/decryption on desktops and few seconds on\shorten{low-power} IoT devices~\cite{ABE-IoT-performance}.
FAME~\cite{agrawal2017fame} exhibits a constant decryption time (60 ms)\shorten{independent of the number of attributes}, however, encryption time increases linearly with the number of attributes.

\shorten{
\textit{Predicate encryption (PE)}~\cite{shi, boneh2007conjunctive} is a paradigm of 
encryption that supports searching on encrypted data with predicate evaluation. 
Hence, it provides the cryptographic means for selective access to encrypted data.
Shi et al.~\cite{shi} propose a PE scheme for range queries over encrypted data, 
where at encryption time, data is associated with a set of attributes. 
It is only possible to decrypt data with attributes in the range supported by the given key.
This scheme is in principle similar to ours,
however, the underlying crypto primitives used in this construction are different, i.e., paring-crypto (Anonymous 
Identity-Based Encryption~\cite{shi}) 
which incurs high overhead in our setting, i.e., large stream data.
}

\shorten{
\fakeparagraph{Key-assignment schemes.}
Our key derivation function resembles key-assignment schemes for hierarchical access control~\cite{crampton2006key}.
Key assignment schemes consider a central key publisher and multiple clients.
While the publisher distributes keys to authorized clients, the individual keys allow a client to access only data according to a hierarchical policy.
Different constructions are introduced for hierarchical and interval-based access control~\cite{atallah2009dynamic, atallah2007incorporating, crampton2006key, crampton2011practical}.
Broadcast encryption~\cite{kogan2006practical, abdalla2000key, wool2000key} (e.g., to protect digital content distribution systems~\cite{kogan2006practical}) also borrows from hierarchical key-assignment for key management.
Similarly, hash-chain-based constructions~\cite{hu2005efficient, briscoe1999marks, liu2003efficient} enable efficient key management in broadcast services.
Key evolution in secure messaging~\cite{unger2015sok, bellare2017ratcheted}, such as double-ratchet, also partially builds on hash-based constructions to provide forward and backward secrecy.
While \oursystem's key derivation scheme builds on these insights, we build a compound key management design consisting of a time-encoded tree structure and dual-key-regression. 
We retrofit our design for time-series data in a decentralized setting without a central entity governing all encryption keys.
}

The notion of \textit{time-encoded keys} in our  access control is similar to Time-Specific Encryption (TSE)~\cite{paterson2010time, cathalo2005efficient}. 
TSE assigns objects to temporal intervals and for each time instance a unique key is generated.
Our scheme differs from TSE in that no central trusted time server is required for the generation and broadcast of epoch keys.
In \oursystem, each data source generates the data encryption keys per epoch locally,
and key distribution is handled over \oursystem's decentralized network.

\fakeparagraph{Distributed Authorization.}
Current distributed authorization protocols, such as OAuth2~\cite{oauth} and 
Macaroons~\cite{macaroons}, suffer from several limitations, as highlighted in \shorten{\rsec{intro} and}\rsec{design:privacy}.

Signature-based schemes (e.g., public-key certificates~\cite{blaze1996decentralized, SPKI}) 
require means for distributing public keys for verification.
Today, conventional approaches to attest to public keys are to rely on internal key servers or at the Internet-scale, hierarchical network of certificate authorities (CA) issuing X.509 certificates,
which come with their weaknesses~\cite{matsumoto2017ikp},
(e.g., Symantec's issuance of unauthorized certificates for Google~\cite{symantec-google}, lack of support for non-domain identities).
Alternative public-key based approaches,
e.g., SPKI/SDSI~\cite{SPKI} and follow-up schemes~\cite{Vanadium},
eliminate the need for complex X.509 public-key infrastructure and CAs. %
However, these schemes are either based on the idea of local names and suitable for deployments under a single 
administrative domain (e.g., smart home) or build upon an organically growing trust model~\cite{pgp} (e.g., PGP's web of trust).
While the key idea of signature-based schemes underpins \oursystem, 
our system neither suffers from certificate-chain discovery nor requires a complex certificate infrastructure (\rsec{design:blockchain}). 
\oursystem's current prototype supports pseudonyms and can be extended with a publicly-auditable directory of keys and identity proofs, such as Keybase,
which maps digital identities (e.g., Twitter) to public keys in a verifiable manner~\cite{Keybase}.

\fakeparagraph{Blockchain-based Systems.}
Decentralized blockchain-based applications (i.e., without trusted intermediaries) beyond cryptocurrencies have gained more 
attention in recent years. 
Example applications include; 
medical data access~\cite{Medrec},
IoT device commissioning and management~\cite{ChainAnchor},
financial auditing~\cite{zkLedger},
name and identity management~\cite{Blockstack, azouvi2017secure},
software-update transparency and verifiability~\cite{chainiac}
and preventing unauthorized certificate issuance~\cite{matsumoto2017ikp}.  %
Closest to our work are;
Enigma~\cite{Enigma, Enigmafull} which envisions a decentralized personal data management and secure multi-party computation platform for multilateral sharing.
They use a single data encryption key among the sharing parties (i.e., no fine-grained crypto-based access) and require blockchain transactions for each read/write request (i.e., limited scalability).
Calypso~\cite{bryan-secret} introduces on-chain encrypted secrets, with associated access policies.
A set of trustees collectively enforces the policies via threshold encryption and distributed key generation,
which ranges for each key access request from 0.2 to 8~s, depending on the number of trustees.
None of the above systems addresses the challenge of fine-grained access control for encrypted time-series data.
Moreover, our design mimics the flow of authorization services in production,
so that \oursystem can seamlessly be integrated to support current services,
as we show through deployments of several case-studies (e.g., serverless computing, \rsec{eval:system}).

\vspace{-5pt}
\section{Conclusion}
\lsec{conclusion}
\vspace{-5pt}

This paper introduces \oursystem, a decentralized access control system that enables secure, selective, and flexible access control that empowers users with full control of their data. 
With \oursystem we present a design that marries a decentralized authorization service and a novel encryption-based access control scheme tailored for time-series data. %
Our prototype implementation and experimental results show the feasibility and applicability of \oursystem as a decentralized authorization service for end-to-end encrypted data streams.

\section{Acknowledgments}
We thank our shepherd Ariel Feldman, the anonymous reviewers, Alexander Viand, Dinesh Bharadia, and Friedemann Mattern for their valuable feedback. 
We thank Simon Duquennoy for his valuable input on earlier versions of this paper.
This work was supported in part by the Swiss National Science Foundation Ambizione Grant, VMware, Intel, and the National Science Foundation under Grant No.1553747.

\vspace{-5pt}
\balance
{\footnotesize 
\bibliographystyle{plain}
\bibliography{sigproc}  %
}

\appendix
\section{\blue{Crypto-based Access Control}}
\vspace{-3pt}
\lsec{proof}
\shorten{Droplet serializes a data stream into a series of consecutive data chunks. 
Each data chunk is encrypted with a unique symmetric key. 
To enable fine-grained access control based on ranges of keys and efficient key sharing, Droplet introduces a hybrid key managed scheme by leveraging particular types of key regression techniques.
We use a \textit{key derivation tree} for generating data encryption keys.
For sharing of past intervals, the tree construction is used, and for sharing in subscription mode, we leverage \textit{dual-key regression}.
In the following, we formally define our dual-key regression and its properties. %
}

\shorten{In the following, we formally define our \textit{key derivation tree} and show that the derived keys are indistinguishable from a random key for a polynomial-time adversary.}
\shorten{In our constructions, we make use of a Pseudorandom Generator (PRG) defined as follows. 

\shorten{\fakeparagraph{Pseudorandom Function (PRF).} 
A function $F: \{0,1\}^\lambda \times \{0,1\}^n \rightarrow \{0,1\}^m$ is a PRF, if there is no probabilistic polynomial-time (PTT) distinguisher, which can distinguish $F_k(x) = F (k, x)$ from a random function drawn from $\{f: \{0,1\}^n \rightarrow \{0,1\}^m\}$ with non-negligible probability in $\lambda$ where $k$ is drawn uniformly at random from $\{0,1\}^\lambda$~\cite{goldreichconstuction}. }
\shorten{
\fakeparagraph{Pseudorandom Generator (PRG).} 
$G: \{0,1\}^n \rightarrow \{0,1\}^m$ is a pseudorandom generator, if $m>n$ and no probabilistic polynomial-time (PTT) distinguisher can distinguish the output $G(x)$ from a uniform choice $r \in \{0,1\}^m$ with non-negligible probability~\cite{goldreichconstuction}.

\shorten{\fakeparagraph{Goldreich-Goldwasser-Micali Construction.} 
The Goldreich-Goldwasser-Micali construction shows how to construct a PRF from a PRG~\cite{goldreichconstuction}. 
Given a PRG $G: \{0,1\}^s \rightarrow \{0,1\}^{2s}$ and $G(x) = G_0(x)||G_1(x)$ both of length $s$, a PRF $F: \{0,1\}^\lambda \times \{0,1\}^n \rightarrow \{0,1\}^s$ can be constructed as follows.
\begin{equation} 
F(k, x=x_1, x_2...x_n) = G_{x_n}(...(G_{x_2}(G_{x_1}(k))))
\end{equation}
Given $G$ is a pseudorandom generator then the above construction is a pseudorandom function.  \newline } }

{
\subsection{\blue{Key Derivation Tree}}
\todo{same text used in timecrypt, make it shorter, and refer to timecrypt}
\lsec{kdt}
In \oursystem, we use a tree-based key derivation function $T$, which allows for efficient access control and sharing of ranges of keys.   
We use a similar construction as presented in~\cite{treekeyreg}. 
We define the function $TreeKD$, which derives keys based on a binary tree with height $h$.
Keys are derived from the $2^h$ leaf-nodes of the tree. Each node in the tree has a unique label $l$ in $\{0,1\}^*$ and an associated \textit{tree-key} in $\{0,1\}^\lambda$.
We define the label of each node in the following manner.
The root node has the label $\epsilon$, the empty-string, 
whereas the left and right children of a node with label $l$ have the labels $l||0$ and $l||1$, respectively. 
Hence, the leaf nodes $ln$ of the tree are indexed by their label $x$ where $x \in \{0,1\}^h$. 
We denote a leaf node with label $x$ as $ln_{x}$.

Each node in the tree with label $l$ has a \textit{tree-key} $z_l$.
The root node of the tree has a randomly chosen \textit{tree-key} $z_\epsilon$ in $\{0,1\}^\lambda$, which corresponds to the input key of the key derivation function.
To derive the keys for the children of a node, a pseudorandom generator $G : \{0,1\}^\lambda \rightarrow \{0,1\}^{2\lambda}$ is used. 
Let $G_0$ and $G_1$ be defined as $G(k) = G_0(k)||G_1(k)$, where $|G_0(k)| = |G_1(k)| = |\lambda|$ and $k \in \{0,1\}^\lambda$. 
The left and right child of a node with \textit{tree-key} $z_l$ are computed as $z_{l||0}=G_0(z_l)$ and $z_{l||1}=G_1(z_l)$.
Hence, the \textit{tree-key} $z_l$ of a leaf-node $ln_{l}$ is constructed as follows:
 \begin{equation}
\lequ{eq:keyderleaf}
 z_l=G_{l_{h}}(...(G_{l_2}(G_{l_1}(z_{\epsilon})))) 
 \end{equation}
 Note that if a \textit{tree-key} $z_l$ is revealed, it is easy to compute the \textit{tree-keys} of its children, but two children \textit{tree-keys} do not reveal any information about the parent \textit{tree-key}.
 This preimage-resistance property allows for access control~\cite{treekeyreg}.
 
 To derive the $t$-th key, the function $TreeKD(k, t)$ computes the \textit{tree-key} $z_t$ of the leaf-node $ln_t$ with root-node key $z_\epsilon = k$ and outputs $z_t$.
 Hence, the function $TreeKD(k, t): \{0,1\}^\lambda \times \{0,1\}^h \rightarrow \{0,1\}^\lambda$ is defined as
 \begin{equation}
    \lequ{eq:keyder}
    TreeKD(k, t) = G_{t_{h}}(...(G_{t_2}(G_{t_1}(k)))) 
\end{equation}
where $t=t_1,t_2,..,t_{h}$. \newline

\fakeparagraph{Theorem 1.} \textit{$TreeKD$ is a pseudorandom function.}
 
\textit{Proof.} This directly follows from the definition of the Goldreich-Goldwasser-Micali construction because $TreeKD$ has an identical definition.\newline
Hence, all the outputs of $TreeKD$ appear to be random for a PPT adversary. \newline
}}

\subsection{Dual-Key Regression}
\lsec{dkr}

A key regression scheme~\cite{fu:keyregression} enables the efficient sharing of past keys.
If an entity is in possession of the key regression state $s_i$, the entity can derive all keys $k_j$ with $j \leq i$ for $i \in \{0, 1, ..., n\}$.
However, the entity cannot infer any information about the keys $k_l$ with $l > i$. 

In our constructions, we make use of a Pseudorandom Generator (PRG) defined as follows. 

\fakeparagraph{Pseudorandom Generator (PRG).} 
$G: \{0,1\}^n \rightarrow \{0,1\}^m$ is a pseudorandom generator, if $m>n$ and no probabilistic polynomial-time (PTT) distinguisher can distinguish the output $G(x)$ from a uniform choice $r \in \{0,1\}^m$ with non-negligible probability~\cite{goldreichconstuction}.

Using a pseudorandom generator $G: \{0,1\}^\lambda \rightarrow \{0,1\}^{\lambda + l}$, a client constructs a key regression scheme as follows.
First, the client generates all the possible states $s_i$ $0 \leq i \leq n$ in reverse order from an initially randomly chosen seed $s_n$.
The seed $s_{i-1}$ is computed as the first $\lambda$ bits of the output of $G(s_i)$. 
To derive key $k_i$ from the corresponding state $s_i$, the client computes $G(s_i)$ and takes the last $l$ bits (i.e., applies the key derivation function). 
For sharing the keys to the $i$-th key, the client shares state $s_i$ with the other entity.
With state $s_i$, the entity can compute all pervious states $s_x$ with $0 \leq x \leq i$ by applying the pseudorandom generator function $G$.
Because of the one-way property of $G$ the client is not able to compute or infer any information about $s_{j+1}$ or any $s_x$ with $x > j$. 
Since the entity owns states $\{s_0, ..., s_i\}$, the entity can derive the keys $\{k_0, ..., k_i\}$ with the key derivation function. 

The key regression scheme based on a single series of states has the drawback that given the current state $s_i$ an entity can compute all the previous states and keys.
Hence, a client is not able to define a lower bound to restrict access on past keys (e.g., $k_j$, $low \leq j \leq cur$).
To address this problem, we combine two sequences of states to derive the keys, similar to~\cite{briscoe1999marks}.
We denote the $i$-th state of the first sequence as $s_{1, i}$ and the second sequence as $s_{2, i}$ for $i \in \{0, \dots, n\}$ where $n + 1$ is the length of each sequence.

In the bootstrapping phase, the client generates the states $s_{1,i}$ as previously from a randomly chosen seed $s_{1,n}$ and computes the other states $s_{1, i-1} = MSB_\lambda(G(s_{1,i}))$ where $MSB_\lambda$ denotes the mapping to the $\lambda$ least significant bits of the input.
The second sequence is generated from the opposite direction to enable a lower restriction level.
The second sequence starts with the random seed $s_{2,0}$ and the corresponding next state is computed as $s_{2,i+1} = MSB_\lambda(G(s_{2, i}))$.
To derive the key $k_j$ where  $j \in \{0, \dots, n\}$, the states $s_{1,j}$ and $s_{2,j}$ serve as an input to the key derivation function which is defined as $k_j=LSB_l(G(s_{1,j} \, xor \, s_{2,j}))$ where $LSB_l$ denotes the mapping to the $l$ most significant bits of the input.
If an entity is in possession of states $s_{1,i}$ and $s_{2,j}$ where $0 \leq j < i \leq n$, it can compute the states $\{s_{1,0}, s_{1,1}, \dots, s_{1,i}\}$ and $\{s_{2,j}, s_{2,j+1}, \dots, s_{2,n}\}$ with $G$.
Since pairs of states are required for deriving the keys, the entity can only compute the keys for which it possesses the corresponding state pairs.
Considering the states computed above, the entity knows the state pairs $\{(s_{1,j}, s_{2,j}), (s_{1,j+1}, s_{2,j+1}) \dots (s_{1,i}, s_{2,i})\}$ and can compute $\{k_j, k_{j+1}, ..., k_i\}$ but no other keys.
Therefore, dual key regression can restrict access based on ranges of keys by sharing the corresponding state of each state sequence. %

\subsection{\blue{Key Derivation Tree}}
\lsec{desc-keyder}
\blue{\oursystem's key-derivation tree is based on the Goldreich-Goldwasser-Micali (GGM) construction~\cite{goldreichconstuction}.
The GGM construction is a binary tree of height $h$ where each node contains a unique binary label $v$ and an associated key $k'$.
The label of a node encodes the path from the root to the current node where the label of the left child is encoded as $v||0$ and the right child as $v||1$.
The key of a node is computed based on the label $v=v_1,v_2,..,v_{l}$ as $G_{v_{l}}(...(G_{v_2}(G_{v_1}(k'))))$ where $G(k') = G_0(k')||G_1(k')$ is a PRG.
The GGM tree is a construction that builds a pseudorandom function (PRF)~\cite{goldreichconstuction}.
The PRF $T$ takes as an input a master key $k$ and a leaf label $v$ and outputs a key $T(k,v) = k_v$.
In GGM, $k$ is the key of the root node, $v$ the label of a leaf node, and the output $k_v$ the key associated with the leaf node with label $v$.
In \oursystem, the keystream for encryption is derived using $T$, which leads to the keystream $\{T(k, 0), T(k, 1),..., T(k, 2^h-1)\}$.}

\blue{
To enable access control on the output keys, $T$ offers the following additional algorithms:
}

\begin{itemize}
    \item\blue{ $\text{\textbf{T.constrain}}(k,S)$ takes as an input the master secret of the root node $k$ and a set of labels of leaf nodes $S$. The algorithm outputs a set of constrained keys $k_S$ that contains the keys from the inner-nodes.  
    These inner-node keys are selected so that they facilitate the computation of the keys of the nodes with labels in S but no other leaf node keys.}
    \item \blue{$\text{\textbf{T.eval}}(k_S, v)$ takes as an input the set of constrained keys $k_S$ and a label $v$ of a leaf node. The algorithm outputs the leaf node key $k_v$ if $v \in S$ else outputs $\bot$.}
\end{itemize}

\blue{
With the two additional algorithms for access control, the construction is a constrained PRF~\cite{boneh2013constrained}.
For the detailed security analysis, we refer to~\cite{boneh2013constrained}.
}

\section{Dual-Key Stealth Addresses}
\lsec{app:stealth}
To protect the privacy of access permissions, \oursystem employs dual-key stealth addresses~\cite{stealth}.
Let us consider the case of a data owner Alice giving access permission to a subscriber Bob.
Bob has initially constructed and published his dual public keys (B, V):  $B=bG$ and $V=vG$, 
with G as the elliptic curve group generator and the private keys b and v.
Alice constructs a new address $P$ using Bob's stealth addresses by using a hashing function $H$, and generating a random salt $r$:
\begin{equation} P = H(rV)G + B 
\end{equation}

Alice embeds the tuple $(P, R)$ in the access permissions, with $R = rG$ ($r$ is protected and not recoverable from $R$).
Only Bob can claim the address $P$, as he is the only one capable of recovering the private key $x$, such that $P = xG$, as follows:
\begin{equation}
\lequ{eq:new-key}
x :=  H(vR) + b
 \end{equation}

Hence, he can prove (e.g., with a signature) to the storage node that he is the rightful principal.
Note that guessing $x$, given $G$ and $P$, is equivalent to solving the elliptic curve discrete log problem, which is computationally intractable for large integers.
The correctness of $x$ from~\req{eq:new-key} can be shown as:
\begin{equation} 
\begin{split}
xG = (H(vR)+b)G = H(vR)G+bG =\\
H(vrG)G+B = H(rvG)G+B = H(rV)G+B = P %
\end{split}
\end{equation}
Except Alice and Bob no other party can learn that $P$ is associated with Bob's stealth addresses.
Moreover, the randomness $r$ in the address generation ensures the uniqueness and unlinkability of new addresses.
Bob discloses the private viewer key $v$ to the auditor to enable an authorized auditor to audit the sharing. 
The auditor can verify the mapping of the tuple $(P, R)$ to Bob's main key address $B$ as:
\begin{equation} 
\begin{split}
P - H(vR)G = P - H(vrG)G =\\
H(rV)G + B - H(rV)G = B     %
\end{split}
\end{equation}
Note that the auditor is cryptographically prevented from using $v$ to compute Bob's private key $x$.

\section{\blue{Security Guarantees}}
\lsec{sec-guarantees}
\blue{
\oursystem consists of the following entities:
\textit{the data owner, data producer, data consumer, storage node, authorization agent, and decentralized authorization log (a public blockchain)}, as described in~\rsec{design}.
Under the trust assumptions laid out in~\rsec{sec-model}, we now elaborate on the security guarantees of \oursystem.
}

\newcommand{\adv}{\texttt{Adv} }%
\fakeparagraph{Guarantee 1.1} 
\blue{
\textit{An \adv is not able to access or manipulate data chunks except by compromising data producers/consumers.}
\oursystem ensures this by end-to-end encryption.
Each data chunk is encrypted with a fresh key (\rsec{desc-keyder}) on the client-side with AES in GCM mode, which is an authenticated block-cipher, providing confidentiality, integrity, and authenticity guarantees:}
\begin{equation} 
\begin{split}
\text{AES-GCM}.{Enc}(K_i, IV, M_i) \rightarrow C_i \\
\text{AES-GCM}.{Dec}(K_i, IV, C_i) \rightarrow M_i
\end{split}
\end{equation}

\blue{Given the $i-th$ key, IV, and  $i-th$ message, it computes the $i-th$ ciphertext.
Given the $i-th$ key, IV, and $i-th$ ciphertext, it computes the $i-th$ message or fails with an {\sf error}.
}

\blue{For proof of ownership, each chunk is digitally signed:}
\begin{equation} 
\begin{split}
\text{ECDSA.KeyGen}(curve) \rightarrow (PK_{device_{id}} , SK_{device_{id}}) \\
\text{ECDSA.Sign}(SK_{device_{id}} , C_i) \rightarrow Sig_{C_i} \\
\text{ECDSA.Verify}(PK_{device_{id}}, C_i, Sig_{C_i}) \rightarrow (\sf{true}, \sf{false}) \\
\end{split}
\end{equation}

\blue{
After generating the per device private and public ECDSA key pair \shorten{over P-256 elliptic curve}, \oursystem signs the encrypted message and generates the signature, which can be verified given the public key and the ciphertext.
As long as the \texttt{Adv} does not compromise the private key, a polynomial-time \texttt{Adv} cannot forge the signature.}

\fakeparagraph{Guarantee 1.2} 
\blue{
\textit{For streams with strong immutability requirements, an \adv is not able to modify the stream without compromising the authorization log.}
The \adv must control a large threshold of nodes/computing power \shorten{(e.g., 1/3 with the PBFT consensus~\cite{bano2017consensus})} to compromise the authorization log to change a committed hash link.
}

\fakeparagraph{Guarantee 2.1.}
\blue{
\textit{If an \adv compromises data consumers that had access to intervals of a data stream,
the \adv is not able to access any other data than the data the compromised data consumers previously had access.}
Each data chunk in a stream is encrypted with a fresh key $K_i$.
If an \adv compromises a data consumer, the \adv gains access to the subset of the decryption keys which the consumer had access.
Hence, it can only decrypt the data chunks where it possesses the decryption keys.
In \oursystem, %
the keys for encryption are derived with a PRF that is constructed from the key derivation tree $T$ (\rsec{desc-keyder}).
With the master secret $k$, the i-th key is derived as $T(k, i) \rightarrow K_i$. 
Instead of sharing the keys for range $K_i,..., K_j$ individually with a data consumer, \oursystem shares constrained keys $T.constrain(k, S:=\{i,..,j\})\rightarrow k_S  $ (i.e., inner-nodes of the tree).
$K_i,..., K_j$ can be derived as $T.eval(k_S, i),.., T.eval(k_S, j)$ but no other keys.
This guarantees that an \adv in possession of $k_S$ can derive keys outside of the interval $K_i,..., K_j$ with negligible probability.
}

\fakeparagraph{Guarantee 2.2.} 
\blue{
\textit{In addition to Guarantee 2.1, an \adv in control of a compromised data consumer can access data that was previously revoked, if the \adv controls the respective storage node or authorization agent.}
Beyond end-to-end encryption, the storage node enforces access based on the authorization log. 
To retrieve data after revoked access, the \adv must compromise the storage node or authorization agent.}

\fakeparagraph{Guarantee 3.1.} 
\blue{
\textit{An \adv cannot link data permissions of a data consumer from the publicly accessible authorization log unless the \adv compromises the audit key of the data owner.}
Dual-key stealth addresses hide any linkability between the consumer identities included in the access permissions~(\rsec{app:stealth}).
A data consumer proves legitimate access to the storage node via a zero-knowledge proof, where the data consumer proves it controls the private key associated with the public key, whose hash digest is included in the access permission.
}

\fakeparagraph{Guarantee 3.2.} 
\blue{
\textit{An \adv compromising the authorization agent cannot compromise data confidentially nor break the non-linkability from Guarantee 3.1, but it can prevent data availability.} 
An \adv can maliciously give access to encrypted data, which does not impact data confidentiality as data is end-to-end encrypted.
An \adv does not learn anything about the data consumer from their request to access their data, other than that they control the private key corresponding to the public key included in the access permission.
}

\shorten{
\fakeparagraph{Guarantee 3.1.} 
\blue{
\textit{An \adv can only grant/revoke access permissions on behalf of data owners that the \adv controls.}
Each policy update has to be signed by the data owner.}
\red{I don't know what to write here... We just need that the storage is auditable, the sever cannot mix the order of updates, and we need freshness. I.e. the server cannot perform a rollback.}

\fakeparagraph{Guarantee 3.2.} 
\blue{
\textit{Access permissions are publicly auditable and cannot be manipulated by an \adv as long as the \adv does not control the authorization log or all authorization agents that a user contacts.}}
\red{same as above }
}

\end{document}